\documentclass[journal]{IEEEtran}
\usepackage{cite}
\usepackage{algorithm}
\usepackage{algpseudocode}
\usepackage{fancyhdr}
\usepackage{textcomp}
\usepackage{amsmath}
\usepackage{subfigure}
\usepackage{amssymb,amsfonts}
\usepackage{graphicx}
\usepackage{textcomp}
\usepackage{xcolor}
\usepackage{multicol}
\usepackage{multirow}
\usepackage{blindtext}
\usepackage{adjustbox}
\algnewcommand{\Initialize}[1]{%
  \State \textbf{Initialize:}
\parbox[t]{.8\linewidth}{\raggedright #1}
}
\algnewcommand{\Goto}{\textbf{go to}}%

\def\BibTeX{{\rm B\kern-.05em{\sc i\kern-.025em b}\kern-.08em
    T\kern-.1667em\lower.7ex\hbox{E}\kern-.125emX}}

\usepackage{graphicx}

\begin{document}
\title{On-Device Intelligence for 5G RAN: Knowledge Transfer and Federated Learning enabled UE-Centric Traffic Steering}

% \author{
%     \IEEEauthorblockN{Han Zhang\IEEEauthorrefmark{1}, Hao Zhou\IEEEauthorrefmark{1}, Medhat Elsayed\IEEEauthorrefmark{2},Majid Bavand\IEEEauthorrefmark{2}, Raimundas Gaigalas\IEEEauthorrefmark{2}, Yigit
%     Ozcan\IEEEauthorrefmark{2} and Melike Erol-Kantarci\IEEEauthorrefmark{1},\IEEEmembership{Senior Member, IEEE}}
    
%     \IEEEauthorblockA{\IEEEauthorrefmark{1} School of Electrical Engineering and Computer Science, University of Ottawa, Ottawa, Canada}
    
%     \IEEEauthorblockA{\IEEEauthorrefmark{2} Ericsson Inc., Ottawa, Canada}
    
%     \IEEEauthorblockA{\{hzhan363, hzhou098, melike.erolkantarci\}@uottawa.ca, \{medhat.elsayed;, majid.bavand, raimundas.gaigalas, yigit.ozcan\}@ericsson.com}
% }
\author{Han~Zhang,
        Hao~Zhou,
        Medhat~Elsayed,
        Majid~Bavand,
        Raimundas~Gaigalas,
        Yigit~Ozcan,
        and~Melike~Erol-Kantarci,~\IEEEmembership{Senior Member, IEEE}% <-this % stops a space
\thanks{Han Zhang, Hao Zhou and Melike Erol-Kantarci are with the School of Electrical Engineering and Computer Science, University of Ottawa, Ottawa, ON K1N 6N5, Canada (e-mail: hzhan363@uottawa.ca; hzhou098@uottawa.ca; melike.erolkantarci@uottawa.ca).}% <-this % stops a space
\thanks{Medhat Elsayed, Majid Bavand, Raimundas Gaigalas and Yigit Ozcan are with Ericsson, Ottawa, K2K 2V6, Canada(e-mail:
medhat.elsayed@ericsson.com; majid.bavand@ericsson.com; raimundas.
gaigalas@ericsson.com; yigit.ozcan@ericsson.com)
}}% <-this % stops a space

%\author{\IEEEauthorblockN{Han Zhang, Hao Zhou, Medhat Elsayed, Majid Bavand, Raimundas Gaigalas and Melike Erol-Kantarci, \IEEEmembership{Senior Member, IEEE}}}


\maketitle

\thispagestyle{fancy}            %更改plain状态，首页格式设为fancy
\chead{This paper has been accepted by IEEE Transactions on Cognitive Communications and Networking. } 

\renewcommand{\headrulewidth}{1pt}      %把页眉线的宽度设为零，即去掉页眉线
\pagestyle{plain}

\begin{abstract}
Traffic steering (TS) is a promising approach to support various service requirements and enhance transmission reliability by distributing network traffic loads to appropriate base stations (BSs). In conventional cell-centric TS strategies, BSs make TS decisions for all user equipment (UEs) in a centralized manner, which focuses more on the overall performance of the whole cell, disregarding specific requirements of individual UE. 
The flourishing machine learning technologies and evolving UE-centric 5G network architecture have prompted the emergence of new TS technologies. In this paper, we propose a knowledge transfer and federated learning-enabled UE-centric (KT-FLUC) TS framework for highly dynamic 5G radio access networks (RAN). Specifically, first, we propose an attention-weighted group federated learning scheme. 
It enables intelligent UEs to make TS decisions autonomously using local models and observations, and a global model is defined to coordinate local TS decisions and share experiences among UEs.
Secondly, considering the individual UE's limited computation and energy resources, a growing and pruning-based model compression method is introduced, mitigating the computation burden of UEs and reducing the communication overhead of federated learning.
In addition, we propose a Q-value-based knowledge transfer method to initialize newcomer UEs, achieving a jump start for their training efficiency. Finally, the simulations show that our proposed KT-FLUC algorithm can effectively improve the service quality, achieving 65\% and 38\% lower delay and 52\% and 57\% higher throughput compared with cell-based TS and other UE-centric TS strategies, respectively.

\end{abstract}


% \fi
\IEEEpeerreviewmaketitle

\vspace{-10pt}
\section{Introduction}

In recent years, the explosive growth of traffic demand has pushed network operators to continuously improve their management strategies to meet diverse service requirements\cite{b2}.
Traffic steering (TS) is an important network function that distributes network traffic load across multiple transmission paths. It makes user equipment (UE) connected to the best serving base station (BS) and choose best bands when multiple BSs provide coverage in the same geographical area \cite{b1}. 
Meanwhile, multiple radio access technologies (multi-RATs), dual connectivity (DC) and mobility also increase the complexity of TS management \cite{b3}. Therefore, how to manage network traffic efficiently has become a very challenging task for 5G beyond and 6G networks\cite{b3-1}. Conventional TS strategies are usually cell-centric, in which BSs make TS decisions for all the connected UEs. For example, \cite{b4} proposed a centralized load-balancing algorithm to decide handovers by calculating an offset value for users in the same cluster. However, such strategies fail to consider the unique demand of heterogeneous UEs with different traffic types and quality of service (QoS) requirements. Additionally, the conventional centralized method may increase the computation burden on the central controller, and lead to privacy and security concerns.
%Since it is difficult to design individual TS strategies for each UE in a centralized manner, most such strategies use a single criterion, such as traffic load, for all UEs to decide whether to steer traffic. However, such a policy may not be the best choice for heterogeneous UEs with different traffic types and quality of service (QoS) requirements.
%\red{For example, xx proposed an xxx, give a short example here.} \red{However, then explain the issues: such as without considering specific demands of each UE}

As the traffic types diversify, UE-centric TS strategies have appeared and attracted more attention. In a UE-centric TS, each UE selects the appropriate RAT, cell and frequency band that best serves their need. Compared with conventional cell-centric methods, the UE-centric scheme indicates that each UE can make its decision using local observations and experience, and then submit it to the BS for coordination. 
%\red{It guarantees that each UE can make the best decision based on }
%Meanwhile, given the service-based architecture of 5G networks, the selection of network nodes and the design of network functions should be device-specific. 
%This leads to a trend to disrupt the cell-centric architecture and to equip UEs with more robust capabilities to meet various communication needs \cite{b30}.
%\red{Can you explain this a little bit?}
%On the other hand,\red{you have two "on the other hand"} as the storage and computational capabilities of the devices within distributed networks grow, it has become a significant trend to leverage enhanced local resources on each device by transferring intellectual processing ability from the cloud to the devices\cite{b26}. Meanwhile, 5G network adopts a UE-centric access network architecture, which equips UEs with more robust capabilities to meet various communication requirements
Hence, UE-centric networks with on-device intelligence have become a promising technique for TS \cite{b5}-\cite{b29}. The benefits of designing a UE-centric network include: 

(1) The UE-centric framework can mitigate the computation burden on the central controller because some computational tasks are transferred to the UEs.

%In addition, such a UE-centric scheme reduces the communication overhead between the APs and UEs. %If a centralized controller makes the TS decision, 
%\red{In conventional cell-centric scheme,} the controller needs to collect information frequently fromributed UEs, result in high signaling overhead \cite{b31}. By contrast, if UEs locally submit their TS demand, there will be less \red{communication overhead between APs and UEs.} 
%that needs to be exchanged, leading to lower overhead.

(2) The UE-centric framework enables better QoS for diverse UEs because each UE can customize its TS demand and submit local decisions to BSs for coordination.

(3) Compared with conventional cell-centric frameworks, UE-centric schemes can better preserve user privacy because some local observations of UE can be locally kept and will not be shared with the centralized controller.

In addition, by leveraging on-device intelligence and offloading part of the processing computation to the UE side, the UE-centric TS framework also has better scalability compared to cell-centric TS. When the number of UE increases, the sum of on-device intelligence also increases. As a result, the framework is able to accommodate a very large number of UEs. Despite the potential benefits, UE-centric TS requires multiple distributed models rather than a single centralized model. The limited local observations of individual UE and coordination between multiple heterogeneous UEs lead to great complexity for the design and implementation. In addition, a single UE may have limited computational and energy resources for decision-making, increasing the difficulty of network management.   
In recent years, machine learning (ML) technologies have been widely applied in wireless communications to provide automated solutions for high-complexity tasks\cite{b33}.
Especially, deep reinforcement learning (DRL)-based algorithms are proposed to deal with the increasing complexity of network environment. For instance, DRL has been used in \cite{b35} to allocate physical resource blocks and control the power level of UEs, and \cite{b39} applies DRL for joint user association and CU–DU placement under the O-RAN architecture. In \cite{b39-1} and \cite{b39-2}, DRL is applied to dynamic resource allocations for 5G networks. %In \cite{b40}, an actor-critic learning-based algorithm has been used for beam selection and the proposed solution can be deployed as an application in O-RAN. 

%\red{For instance, xx, DRL is used in xxx for..}
%\red{Still keep this for the consistency of the article. Add fl below.}
%\noteblue{On the other hand, federated learning techniques have been widely used to train a robust common model through multiple decentralized devices.}
%In this context, we applied DRL techniques to improve the UE-centric TS framework. 
% \red{focus on the federated learning, not DRL } 
%Some typicathis informationesigning a UE-centric framework should be taken into consideration:

%some problems may be caused by directly applying UE-centric DRL to TS. 
However, there are challenges in implementing UE-centric ML algorithms. Since each UE is considered an independent agent without direct information exchange with other UEs, it can only obtain limited information from local observations. 
%and accumulate a small amount of experience in a given period. 
Moreover, decentralized models may take long iterations to adapt to environmental changes, leading to a relatively long exploration time and slow convergence. In addition, every time a new UE is admitted to the network, it must start learning from the beginning, and this will degrade the overall network performance. To this end, we adopt an attention-weighted federated learning technique and use a global server to coordinate local models. In particular, the decentralized UEs upload the parameters of their local models to the global server. According to the performance of local models, an attention weight is calculated for each UE using a sigmoid function. The global server aggregates the local models into a global model by calculating the attention weights of the local models. Then the global parameters will be returned to the local model as feedback for model updates. Compared with conventional federated learning, the attention mechanism can improve model aggregation and obtain a better global model. Considering the advantages of accelerated convergence, federated learning can be widely used in distributed learning-based network control applications such as distributed resource allocation and edge caching.

%This enables experience sharing among UEs without exchanging local data. %\red{Two more sentences to explain your FL model}

Meanwhile, in a dynamic network environment, existing users may leave the network, and new users will join the network without any intelligent experience. Then these inexperienced new users may degrade the system performance, since they have no knowledge of TS strategies.
To this end, a knowledge transfer approach is incorporated into the proposed framework. Specifically, knowledge acquired by existing users will be transferred to new users and give newcomers a good initialization for a quick start \cite{b14}. It will reduce the training efforts of new users, and mitigate performance degradation caused by inexperienced new UEs.

%\red{We need more text to explain knowledge transfer, such as motivations, benefits, techniques. Imitating the next paragraph about model-compression.}

In addition, considering the large size and complex structure of ML models in most cases, training ML algorithms are supposed to be computation-intensive \cite{b32}. A UE may have limited computational resources for algorithm training and neural network updating, especially when running multiple applications simultaneously. Moreover, these computation-intensive tasks incur extra power consumption for the UE \cite{b38}. As a result, the deployed ML models are expected to be computationally resource-efficient. To address this issue, we proposed a growing and pruning-based model compression technique. By gradually splitting the most competitive neurons and eliminating the weakest neurons, we can estimate the smallest model size without degrading the TS performance. 

Finally, inspired by these benefits and challenges, we propose a knowledge transfer and federated learning-enabled UE-centric (KT-FLUC) TS framework in conjunction with the concept of on-device intelligence. Different from the conventional TS schemes, each UE can make individual TS strategies by training ML models on the UE side. Through this structural innovation, our proposed framework has advantages in scalability, performance, and privacy protection. The main contributions are:

(1) We use several innovative techniques to solve the challenges of implementing UE-centric TS. We propose a unique transfer learning approach of providing expert q-values by building local expert models, which have been barely used in existing studies. We also combine grouping techniques and attention mechanisms with the conventional FL techniques to make it more suitable for our proposed scenario. We considered a time-
variant network in which new UEs may arrive and leave the network dynamically \cite{b16}.

(2) We design a KT-FLUC framework for UE-centric TS. Different from existing frameworks that simply bring different techniques together, in the proposed framework, these innovative techniques are inter-playing and mutually enhancing each other as a whole. In this work, the transfer learning technique makes the level of training of participants in federated learning relatively consistent. This process facilitates the cooperation between participants and thus contributes to FL. On the other hand, federated learning provides a suitable expert model for transfer learning through global aggregation and opens up potential applications for transfer learning. Working in conjunction with each other, each of these technologies can be made more useful in the designated scenario.

(3) We also propose a novel growing and pruning method to compress the DRL model and determine the optimal neural
network (NN) size for TS. In this method, the neuron number of hidden layers of each UE is firstly increased and then decreased. The network performance under different model sizes is recorded, and the optimal number of hidden neurons can be established according to the performance decay rate with varying degrees of model compression. This method can effectively reduce the computation burden and power consumption on devices. To perform model compression more efficiently, we also propose a novel neuron splitting method, and this is the first time we apply model compression to a wireless communication application.

The rest of this work is organized as follows. Section \ref{s1} introduces the related work. Section \ref{s2} shows the system model and problem formulation. Section \ref{s3} explains the KT-FLUC framework and the model compression. The simulations are shown in Section \ref{s5}, and Section \ref{s6} concludes this work.

\vspace{-10pt}
\section{Related Work}\label{s1}
TS has been widely investigated under cell-centric frameworks.
% Among these works, \cite{b4}, \cite{b6} - \cite{b9} adopt cell-centric frameworks while \cite{b5}, \cite{b10} - \cite{b13} applicationsntric frameworks.
In \cite{b4}, a centralized DRL framework is designed to decide the individual cell offset to control the handover of UEs. \cite{b6} studied how to offload a mobile UE from macro cells to small cells and derived optimal offloading strategies with resource partitioning. In \cite{b8}, a centralized radio access network (RAN) management module is designed to make TS decisions according to the traffic load of different RAT. In \cite{b9}, a deep Q-network (DQN) is used to make traffic steering decisions with multiple RATs in a dynamic environment. In \cite{b41}, a load-aware traffic steering algorithm is performed based on centralized hierarchical reinforcement learning.

These studies demonstrate that cell-centric TS can improve network efficiency and average network performance. However, in most existing works, there will be considerable overhead between the centralized controller and distributed UE. The centralized controller must process a large amount of information which will lead to a heavy computation burden. On the other hand, in a centralized architecture, the information submitted by the UEs may contain private information, such as traffic profiles that may infer UE locations and so on. The data exchange between the UEs and the controller makes the system more vulnerable to eavesdropping or attacks. As a result, privacy and security concerns are involved. Moreover, most of these frameworks fail to provide individual TS strategies for each UE, which may not satisfy the varying requirements of different UEs. 

Meanwhile, some existing works investigate UE-centric frameworks with ML techniques. \cite{b5} studies how to design a UE-centric network selection framework in ultra-dense heterogeneous networks by choosing the optimal RAT according to the link state prediction. \cite{b10} provides a fully distributed TS approach running on distributed devices and optimizes each UE's received signal strength and energy consumption. In \cite{b11}, distributed UEs choose their cooperative BS according to the received signal interference noise ratio (SINR) and predicted load of BS. \cite{b13} designed a UE-based distributed TS mechanism between Wi-Fi networks and LTE with Q-learning. Although the above works adopt UE-centric frameworks and some use ML techniques such as federated learning, our work differs from existing works in three aspects. Firstly, most existing works simply deploy models on the UE side and fail to consider the limited resource of UEs. As a result, they may impose extra computation burdens on devices. Secondly, the collaboration between decentralized UEs is rarely considered in previous works, leading to low efficiency when applying the algorithm in practice. In addition, these works consider a static network environment with a fixed number of users, which is unrealistic, but our work includes a dynamic environment where UEs may join and leave the network. This work combines the concepts of reinforcement learning, federated learning, knowledge transfer, and model compression for the first time, merging the advantages of different techniques to collaboratively set up a UE-centric framework for TS.

\vspace{-10pt}
\section{System Model and Problem Formulation}\label{s2}

\subsection{Architecture}
\begin{figure}[t]
\centerline{\includegraphics[width=3.2in]{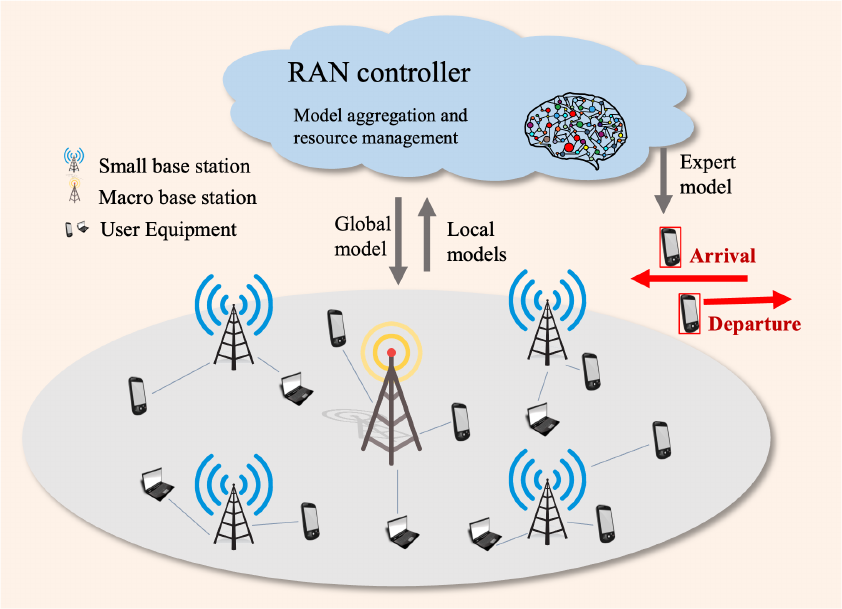}}
\caption{Dynamic wireless networks with one MBS and several SBSs.}
\label{fig1}
\vspace{-10pt}
\end{figure}

The proposed TS system model of this work is shown in Fig.~\ref{fig1}. We consider a network with two coexisting RATs, 5G and long-term evolution (LTE), with DC for downlink transmission. We assume the system has $M$ UEs and $N$ BSs. One BS is a macro BS (MBS) equipped with the LTE technology, and the remaining $(N-1)$ are small BSs (SBSs) equipped with the 5G technology. Each UE can choose a single BS at one time for better service. BSs generate interference with each other due to frequency reuse. Meanwhile, dynamic UE arrival and departure are involved, and we assume UEs arrive at the cell with a fixed file length to be transmitted\cite{b36}.

We assume UEs have two types of services, guaranteed bit rate (GBR) service and non-GBR service, which have different quality of service (QoS) requirements \cite{b37}. Each UE either runs GBR traffic or non-GBR traffic. Most GBR services are sensitive to transmission delay, while non-GBR services require higher throughput \cite{b17}. So, the optimization goal of this TS framework is to minimize the packet delay of GBR services and maximize the throughput of non-GBR services. 
\vspace{-10pt}
\subsection{Communication Model}

The communication delay consists of two components, the queuing delay and the transmission delay:
\begin{equation}
d_{m} = d^{que}_{m} + d^{tx}_{m},\label{eq1}
\end{equation}
where $d^{que}_{m}$ is the queuing delay of the last packet transmitted to the user $m$. It refers to the delay incurred while the packet waits in the BS before being transmitted. $d^{tx}_{m}$ is the transmission delay of the packet. It refers to the time pushing the bits onto the link, which is given as:
\begin{equation}
d^{tx}_{m} = \frac{L_{m}}{C_{n,m}},\label{eq2}
\end{equation}
where $L_{m}$ denotes the packet length and $C_{n,m}$ denotes the link capacity between the $n^{th}$ BS and the $m^{th}$ UE. According to Shannon's theorem, the link capacity is influenced by the signal interference noise ratio (SINR) between the BS and the UE, which can be formulated as:
\begin{equation}
C_{n,m} = \sum_{r\in R_{n}} B_{r} log_{2}(1+\eta_{n,r,m}),\label{eq3}
\end{equation}
where $R_{n}$ denotes the set of all the available resource blocks of the $n^{th}$ BS and $B_{r}$ denotes the available bandwidth of the $r^{th}$ resource block. $\eta_{n,r,m}$ denotes the SINR between the transmission link between the $n^{th}$ BS and the $m^{th}$ UE with the $r^{th}$ resource block. It can be given as:
\begin{equation}
\eta_{n,r,m}=\frac{\alpha_{n,r,m} g_{n,m} P_{n,r}}{\underset{n'\in N, n'\neq n}{\sum}\ \underset{m'\in M_{n'}}{\sum} \alpha_{n',r,m'} g_{n',m} P_{n',r}+B_{r}N_{0}},\label{eq4}
\end{equation}
where $\alpha_{n,r,m}$ is a binary indicator denoting whether the $r^{th}$ physical resource block of the $n^{th}$ BS is allocated to the $m^{th}$ UE. $g_{n,m}$ is the channel gain of the transmission link between the $n^{th}$ BS and the $m^{th}$ UE which is decided by the free space propagation model. $M_n$ denotes the set of UE associated with the $n^{th}$ BS. $P_{n,r}$ denotes the transmission power and $N_{0}$ denotes the power density of the noise. 
%\red{can we add one or two equations that are related to TS? is it possible}

\begin{figure*}[htbp]
\centerline{\includegraphics[width=6.0in]{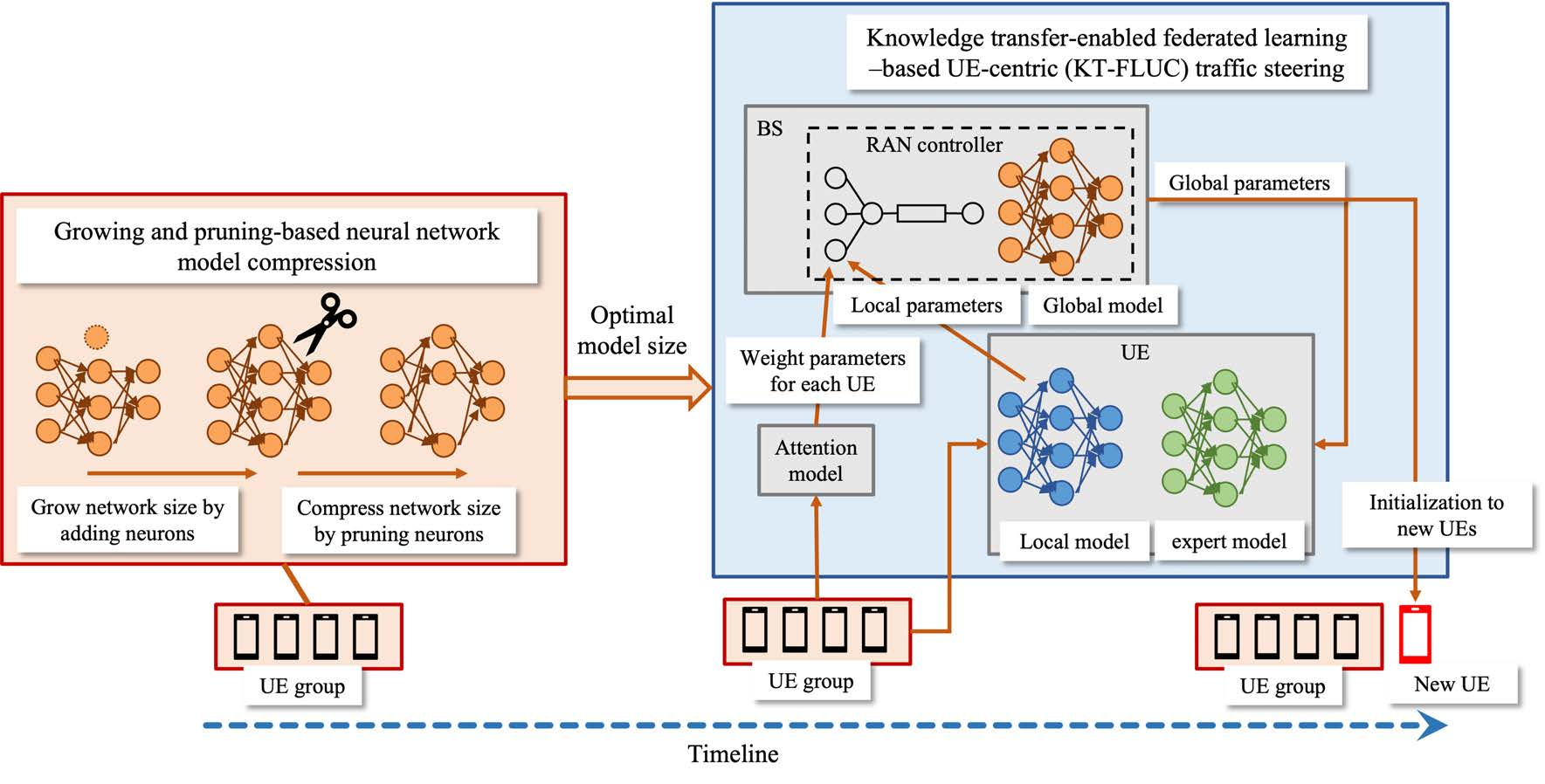}}
\caption{The framework of proposed UE-centric TS method.}
\label{fig2}
\vspace{-10pt}
\end{figure*}
\vspace{-10pt}
\subsection{Problem Formulation}
According to the service type of UEs, we have different optimization objectives for different UEs since they have different QoS requirements. If a UE has a GBR service, the utility of a UE is decided by the delay of the traffic, which is given by Eq (\ref{eq1}). If a UE has a non-GBR service, the utility of a UE is decided by the throughput of the UE, which is decided by the packet arrival and the channel capacity given by Eq (\ref{eq3}). To constrain the utility value between 0 and 1, we use the estimated maximum throughput and delay of UEs, $B_{max}$ and $D_{max}$, for normalization. Furthermore, since we want to minimize the delay time rather than maximize it, we define the utility of GBR traffic as $1-\frac{d_m}{D_{max}}$. Therefore, for the $m^{th}$ UE, the TS problem is formulated to maximize the utility:
\begin{align}
 \underset{\beta_{m,n}}{max}\ &\ \ U_{m}, \label{eq7}\\
s.t. \  &(\ref{eq1})-(\ref{eq2})\nonumber
\\& U_m = \begin{cases}
&\frac{b_m}{B_{max}}, \ \ \ \ \ \ \ \ \ \ \ \ \ \ \ if\ T_m = 0
\\& 1-\frac{d_m}{D_{max}},\ \ \ \ \ \ \ \ \ \ \ if\ T_m = 1
%\\& \frac{1}2{}(1-d_n + tan^{-1}(b_n)), \ \ if\ T_n = 2
\end{cases}\tag{5a}\label{eq7a}
\\& \alpha_{n,r,m} = \{0,1\} \forall n, r\tag{5b}\label{eq7b}
\\&\sum_{m \in M_n}\alpha_{n,r,m}=1\ \ \ \forall r \in R_n, n \in N \tag{5c}\label{eq7f}
\\& M_n = \{m'| \forall \beta_{m',n}=1\}\tag{5d}\label{eq7c}
\\& \beta_{m,n} = \{0,1\}, \forall n\tag{5e}\label{eq7d}
\\& \underset{n \in N}{\Sigma}\beta_{n,m} = 1\tag{5f}\label{eq7e}
%\\& \underset{m \in M_n}{\Sigma}b_{m} < b_{m,n}\tag{7f}\label{eq7f}
\end{align}
where Eq \eqref{eq1}-\eqref{eq2} defines the variable $d_m$ which used in the definition of the utility. Eq \eqref{eq7a} indicates that the optimization goal of each UE is related to its traffic type $T_m$. $T_m = 0$ indicates the $m^{th}$ UE holds non-GBR traffic and the utility is related to its throughput $b_m$. $T_m = 1$ indicates the $m^{th}$ UE holds a GBR traffic, and its reward is related to its delay $d_m$. $B_{max}$ and $D_{max}$ are the estimated maximum throughput and delay of UEs and they are fixed values to normalize the utility of delay and throughput. $\beta_{m,n}$ is a binary indicator denoting whether the $m^{th}$ UE chooses the $n^{th}$ BS. Eq \eqref{eq7b} indicates that a PRB can be assigned or unassigned to a user. Eq \eqref{eq7f} indicates that each physical resource block of a BS can only be allocated to one UE at the same time. Eq \eqref{eq7c} indicates that $\beta_{m,n}$ decides the attached list of BSs and Eq \eqref{eq7e} indicates that each UE can only choose one BS at one time. Given the UE-centric structure of our proposed framework, it is worth noting that the problem formulation in Eq (\ref{eq7}) is defined for one single UE. According to this problem formulation, each UE makes local decisions to maximize its own utility.
%\red{A paragraph to explain that the problem is defined for each UE, it is a UE-centric model, and each UE can make local decisions to maximize its utility. }

As it can be observed in Eq \eqref{eq7}, the proposed problem contains uncertain or stochastic parameters, and the optimal solution may depend on the probability distribution of the parameters. In addition, in the TS scenario, the BS selection of a UE at a given moment affects not only the immediate QoS rewards but also the QoS rewards in the future. We want to focus more on the long-term rewards, and it is difficult to complete by optimization. Therefore, the formulated problem could not be directly solved, and an AI-based solution is proposed.
%\eqref{eq7f} ensures the sum transmission rate of all the attached UEs of a BS will not exceed the maximum transmission rate of the BS.

\vspace{-10pt}
\section{Proposed TS Scheme}\label{s3}
%\red{when we want to introduce a huge section, we will briefly explain what will be covered in this section. }
In this section, we present the proposed KT-FLUC method. First, we provide an overview of the algorithmic framework. Then, we describe the federated reinforcement learning algorithm, model compression method, and knowledge transfer technique we used in KT-FLUC, respectively.
\vspace{-10pt}
\subsection{Algorithm Framework}
Fig.~\ref{fig2} shows the framework of the proposed UE-centric TS method. 
Firstly, we apply the growing and pruning method to decide the optimal network size for the UR intelligence.
In particular, we first grow the NN size by adding neurons, then compress the NN size by pruning neurons. 
Given the appropriate NN size, federated learning and knowledge transfer are cooperatively performed within a KU-FLUC framework. This framework uses an attention-weighted federated learning algorithm to aggregate experience from distributed UEs and generates a global model. A weight parameter is calculated for each UE by an attention model according to the performance of each UE. Then, UEs submit the parameters of their local models to the RAN controller, and the parameters of the global model are updated according to the local parameters and attention weights. After that, the knowledge transfer technique is used to transfer the aggregated experience from the global model to UEs. It is implemented by defining an expert model for each UE. The expert is a copy of the aggregated global model, which runs inside each UE. In the local training process of each UE, only local models are trained, and expert models remain unchanged. Every time the global model is updated, the parameters of expert models will also be updated by parameter replication, and the action selection is processed inside each UE according to the outputs of expert models and local models. 

In addition, for realistic dynamic environments, new UEs may join the network without prior knowledge of intelligent TS. Knowledge transfer will be performed between the global model and the newcomer for initialization. After that, the newcomer will perform local training and join the next round of federated learning.

This KT-FLUC framework uses federated learning to collect and aggregate experience from UEs with similar traffic types and QoS requirements. Knowledge transfer transfers the aggregated experience to existing UEs and newcomers. The expert model of knowledge transfer is obtained during federated learning, which is how these two techniques are connected \cite{b19}. The update of the global model by federated learning is non-real-time, while the TS decisions made by UEs are real-time. 
Under this framework, UEs may leave the system, but their experience will be kept for newcomers. The shared model is constantly updated with the distributed efforts of all the UEs in this network. Despite the high dynamics of the system, the knowledge learned by UEs consistently contributes to the optimization of a shared expert model. In turn, the expert model accelerates the local models' training and enhances system performance.

In the following, we first introduce the federated reinforcement learning-based UE-centric TS algorithm since it serves as the main scheme in this work. Then, we explain how to implement model compression for neural networks, and finally, transfer learning techniques are introduced.

\vspace{-10pt}
% \section{Implementation details}\label{s4}
\subsection{Federated Reinforcement Learning based UE-centric TS}
For UE-centric TS, each UE is treated as an independent agent, including a DQN for decision-making \cite{b20}. The Markov decision process (MDP) of UE-centric TS is defined as:
\begin{itemize}
    \item State: The state of TS of each UE is composed of the received signal strength indicator (RSSI) from each BS, the number of queued packets, the delay of the last arrived packet, and its current attached BS. The RSSI information is accessible to UEs, and it can reveal the state information of the system, including the channel quality and the distance between the UE and BSs. The number of queued packets is related to the QoS of the UE. The delay of the last arrived packet, and the current attached BS indicates whether the UE needs to switch to a new BS. These indicators are important for TS decisions and can easily be observed and collected from the view of a single UE. The state at time $t$ is given as:
    \begin{align}
    S_{m,t} = \{\Gamma_{n,t},L^{queue}_{m,t},d_{m,t},\beta_{m,n,t}|n\in N\}, \label{eq8}
    \end{align}
    where $\Gamma_{n}$ denotes the RSSI from the $n^{th}$ BS and $L^{queue}_m$ denotes the current packet queue length.
    
    \item Action: After observing states from the environment, UE will make local TS decisions. Since TS is to select the best serving BS for each UE, the action of TS of each UE is to choose an optimal BS, which is given as:
    \begin{align}
    A_{m,t} = \{\beta_{m,n,t}|n\in N\},, \label{eq9}
    \end{align}
    where $\beta_{m,n}$ denotes whether the $m^{th}$ UE chooses the $n^{th}$ BS.
    
    \item Reward: The TS reward for each UE is related to their traffic type and the delay and throughput of the traffic, which can be given as:
    \begin{align}
    \ & R_{m,t} = \begin{cases}
    &\frac{b_{m,t}}{B_{max}}, \ \ \ \ \ \ \ \ \ \ if\ T_n = 0.
    \\& 1-\frac{d_{m,t}}{D_{max}}%,
    ,\ \ \ \ \ if\ T_n = 1.
    \end{cases}\label{eq10}
    \end{align}
\end{itemize}

As it can be observed from the MDP definition, the definition of reward has a similar format to the utility, which is the optimization goal of the problem formulation. The action is similar to the control variable of the problem formulation. By doing this, we convert the problem defined in Eq (\ref{eq7}) into a MDP problem, which can then be solved with the DQN algorithm. The optimal gap of the solution is decided by the training effect of the DQN. In an ideal situation, the DQN is able to develop an optimal TS solution with full exploration and accurate prediction. However, it is very difficult for DQN to be trained to perfection in real cases. To reduce the optimal gap, in this work, we also use FL and transfer learning techniques to enhance the training effect of DQN. Also, to limit the number of handovers, the handover action will be selected only if the reward value differential is higher than a given threshold. In this way, the framework will avoid unnecessary handovers.

The core idea of DQN is to use a NN to predict the long-term reward, which is denoted as Q-values, of different actions under given states. The Q-values in each step can be updated according to the Bellman equation: 
\begin{align}
Q(s_{m,t},a_{m,t}) = (1-\alpha)Q(s_{m,t},a_{m,t}) \\ +\alpha(R_{m,t} + \underset{a}{max}\gamma Q(s_{m,t+1},a)),\nonumber\label{eq11}
\end{align}
where $Q(s_{m,t},a_{m,t})$ denotes the long-term reward of choosing action $a$ under the observation of state $s$ at time $t$ of the $m^{th}$ UE. $\gamma$ is the discount factor denoting the importance of distant future reward compared with instant reward, and $\alpha$ is the learning rate denoting the update speed of the NN. The local update of DQN at the $m^{th}$ UE can be formulated as:
\begin{equation}
\begin{split}
\theta_{m,t+1} &= \theta_{m,t} + \alpha[R_{m,t} + \gamma \mathop{max}\limits_{a}Q(s_{m,t+1},a;\theta_{m,t}) \\ & -Q(s_{m,t},a_{m,t};\theta_{m,t})]\nabla Q(s_{m,t},a_{m,t};\theta_{m,t}),\label{eq12}
\end{split}
\end{equation}
where $\theta_{m,t}$ denotes the parameters of the DQN at time $t$. 

\begin{figure}[tbp]
\centerline{\includegraphics[width=3.3in]{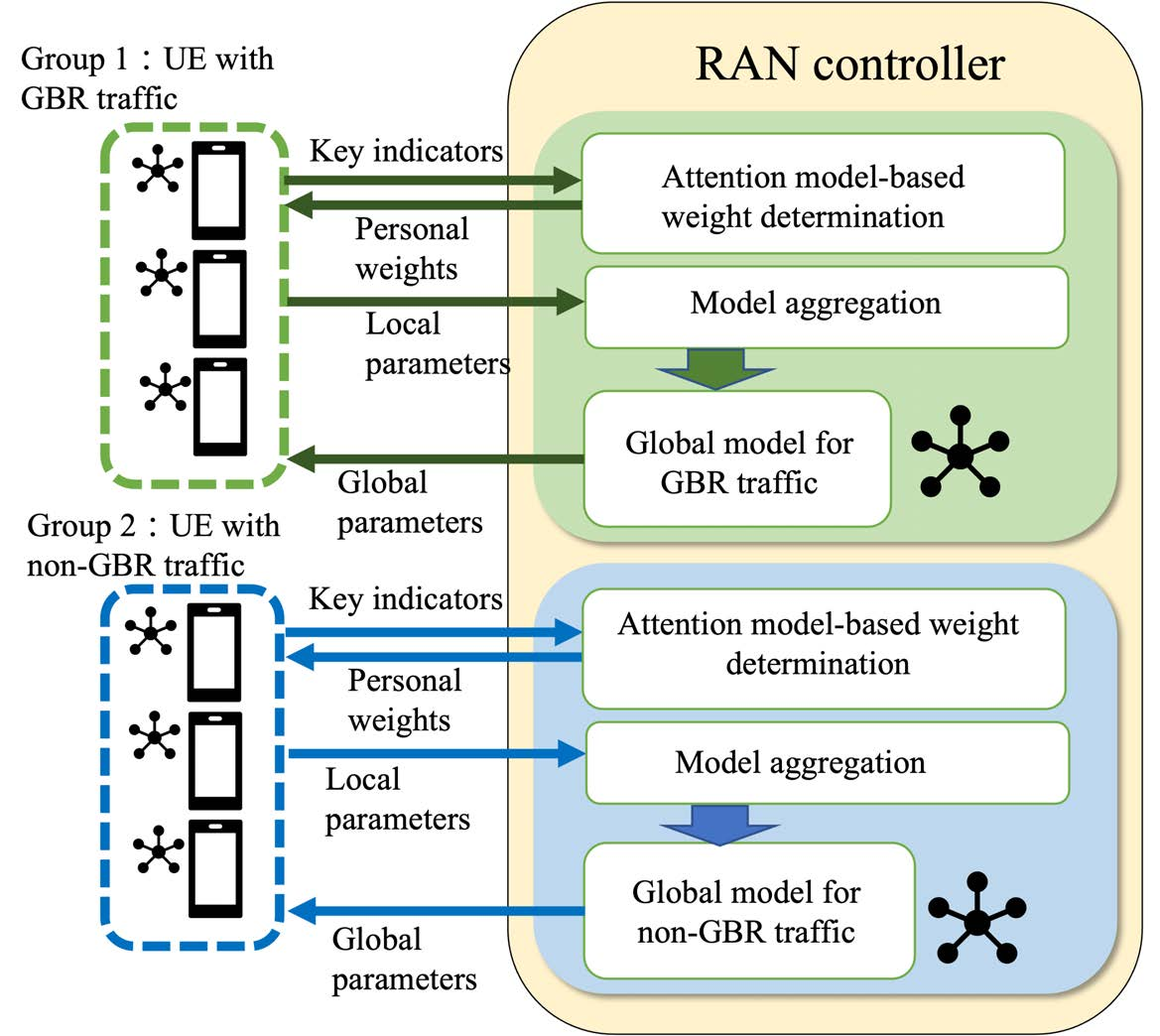}}
\caption{Federated learning process.}
\label{fig4}
\vspace{-5pt}
\end{figure}

% This section introduces the implementation details of the federated learning technique adopted in our proposed TS framework.
The federated learning scheme is shown in Fig.~\ref{fig4}. Here, we adopted a grouped attention-weighted federated learning approach based on FedAvg, and multi-centered federated learning \cite{b15}-\cite{b25}. Considering the heterogeneity between different traffic types, we assume UEs with GBR traffic as the first group and UEs with non-GBR traffic as the second group and enable model federating and experience sharing among groups. Inside each group, an attention weight is calculated and assigned to each UE to represent the contribution of each UE to the global model and promote model aggregation.

\begin{algorithm}[t]
\caption{Local training at each UE}\label{alg:localT}
\begin{algorithmic}[1]
    \State Observe local state.
    \State $S_{m,t} \gets \{\Gamma_{n},L^{queue}_n,d_m,\beta_{m,n}|n\in N\}$.
    \State Get immediate reward $R_{m,t}$.
    \State Record experience $\{S_{m,t+1},S_{m,t},a_{m,t},R_{m,t}\}$.
    \State Sample a minibatch from experience buffer
    \State Update local Q network $\theta$ with $\nabla(L^{local} (\theta_{m,t}))$ via  Equation (\ref{eq12})
    \State Choose actions by $a_{m,t} \gets \underset{a_{m,t}}{argmax}Q(S_{m,t},a_{m,t};\theta_{m,t})$.
    \State Execute action $a_{m,t}$.
\end{algorithmic}
\end{algorithm}

The federated learning process inside each group can be concluded as three main steps. The first step is the local training and updates summarized in Algorithm \ref{alg:localT}. In this step, each UE observes local states, selects actions according to the Q-values generated by local DQN, and uses local experience to train DQN by minimizing the local loss function. The local loss function is defined as the mean squared error, which can be given as:
\begin{align}
Loss_m^{local}(\theta_{m,t}) = \underset{\Psi_m}{\sum}(y_m-R_{m,t}-\\\gamma\underset{a}{\max}{Q(s_{m,t+1},a};\theta_{m,t}))^2,\nonumber
\end{align}
where $Loss_m^{local}$ denotes the local loss function of the $m^{th}$ UE and $\Psi_m$ denotes the memory buffer of the $m^{th}$ UE. $y_m$ denotes the output of the local DQN.

The second step is attention weights determination. In each federating interval, all the UEs submit the parameters of their local DQN to the RAN controller to aggregate a global model. Weight parameters are calculated and assigned for each UE as preparation for aggregation. Since all the local models are trained separately and independently, they are trained at different degrees. To generate a more accurate and generalized global model, we expect well-trained models to contribute more to the global model. In contrast, barely trained models contribute less to the global model. Given this, three key indicators are used to evaluate how well the model has been trained, the average reward, training data size and the achievement rate.

\begin{itemize}
    \item The average reward of a UE is denoted as $\overline{R_m}$ and refers to the average value of system reward defined by Equation (\ref{eq10}) of the local DQN during the interval of two federating processes. It can be seen as an intuitive representation of the DQN model performance.
    \item The training data size of a UE is denoted as ${|\Psi_m|}$ and refers to the amount of training data. We expect UEs with richer experience to account for more weight in federating.
    \item The achievable rate of a UE is denoted as $\Phi_m$ and refers to the frequency that the UE achieves a specific level of latency or throughput requirement. We add this indicator to $\overline{R_m}$ to prevent coincidental large rewards from giving a false representation of model performance.
\end{itemize}    

After obtaining all the key indicators from the UE side, we use a SoftMax function-based attention mechanism to combine these indicators and decide a weight for each local model accordingly:
\begin{align}
    w_m = softmax(\frac{QK_m}{\sqrt{n_k}}), 
\end{align}
where $w_m$ denotes the weight of the $m^{th}$ UE. $n_k$ is the dimension of indicators. $K_m$ is the key indicators and can be formulated as $K_m=[\frac{\overline{R_m}}{\underset{m}{max}\overline{R_m}},\frac{{|\Psi_m|}}{\underset{m}{max}{|\Psi_m|}},\frac{\Phi_m}{\underset{m}{max}\Phi_m}]$. $Q$ is the query and can be formulated as $Q=[1,1,1]$. By defining the key and the query, we expect UEs with indicators closer to the maximum values receive more attention. For instance, if a UE has a higher average reward value and a higher achievable rate, it is considered to be better trained. If it has a larger training data size, the training may considered to be more reliable. Therefore, we firstly pre-process the indicators by finding the maximum $\overline{R_m}$, ${|\Psi_m|}$ and $\Phi_m$ among all the UEs and perform normalization. The weight can then be calculated as:
\begin{align}
w_m &= softmax(\frac{\overline{R_m}'+{|\Psi_m|}'+\Phi_m'}{\sqrt{n_k}})\nonumber
\\&= \frac{\exp{(\frac{\overline{R_m}'+{|\Psi_m|}'+\Phi_m'}{\sqrt{n_k}}})}{\underset{m' \in M}{\sum}\exp{(\frac{\overline{R_{m'}'}'+{|\Psi_{m'}'|}'+\Phi_{m'}'}{\sqrt{n_k}})}},
\end{align}
where $\overline{R_m}'$, ${|\Psi_m|}'$ and $\Phi_m'$ denotes the indicators after normalization. 

Finally, the last step is model aggregation and model update. The parameter of global model can be obtained as the weighted sum of all the local models, which is given as:
\begin{align}
\theta_{g,t+1} = (1-\eta_1)\theta_{g,t} + \eta_1 w_m \sum_{m \in M}\theta_{m,t}, \label{eq24}
\end{align}
where $\theta_{g}$ denotes the parameters of the global model. $\eta_1$ is a parameter denoting the update speed of global model parameters. $\theta_{m}$ denotes the parameters of the $m^{th}$ local model. In this way, the well-trained models will be given larger weights and contribute more to the global model. The barely trained models will be given smaller weights and contribute less to the global model.

Then the local models are updated with the global model, which is given as:
\begin{align}
\theta_{m,t+1} = (1-\eta_2)\theta_{m,t} + \eta_2\theta_{g,t},\label{eq25}
\end{align}
where $\eta_2$ is a parameter denoting the update speed of local model parameters during federating.
The grouped attention-weighted federated reinforcement learning approach is summarized at Algorithm \ref{alg:fl}, where $Eligible_m$ is a binary indicator denoting whether the UE achieves a specific level of latency or throughput requirement and $FedInterval$ denotes the time interval between two times of model aggregations in federated learning.

\begin{algorithm}[t]
\caption{Grouped attention-weighted federated reinforcement learning-based TS}\label{alg:fl}
\begin{algorithmic}[1]
\Initialize{$\theta_{local}$ of each UE, $S^{t-1}$ and $a^t$}
 \While {$TTI < T^{total}$}
    \For{each UE}
    \State Do local training via Algorithm \ref{alg:localT}
    \EndFor
    \If{$mod(TTI,FedInterval)==0$}
        \For{each UE} 
            \State Submit $\theta_{local}$ to RAN controller.
            \State $\overline{R}_{n} \gets \underset{TTI \in FedInterval}R_{TTI}/FedInterval$
            \State $|\Psi_m| \gets length\{buffer\}/T^{total}$
            \State ${\phi_m}\gets \underset{TTI \in FedInterval}{\sum}Eligible_m/FedInterval$
            \State $w_n \gets softmax(\overline{R}_{n},|\Psi_m| ,\phi_m)$
        \EndFor
        \State Update global model according to Equation (\ref{eq24})
        \For{each UE} 
            \State Update local models via Equation (\ref{eq25})
        \EndFor
        \State $TTI \gets TTI + 1$
    \EndIf
\EndWhile
\end{algorithmic}
\end{algorithm}

\vspace{-5pt}
\subsection{Growing and Pruning-based Model Compression}

\begin{figure}[t]
\centerline{\includegraphics[width=3.4in]{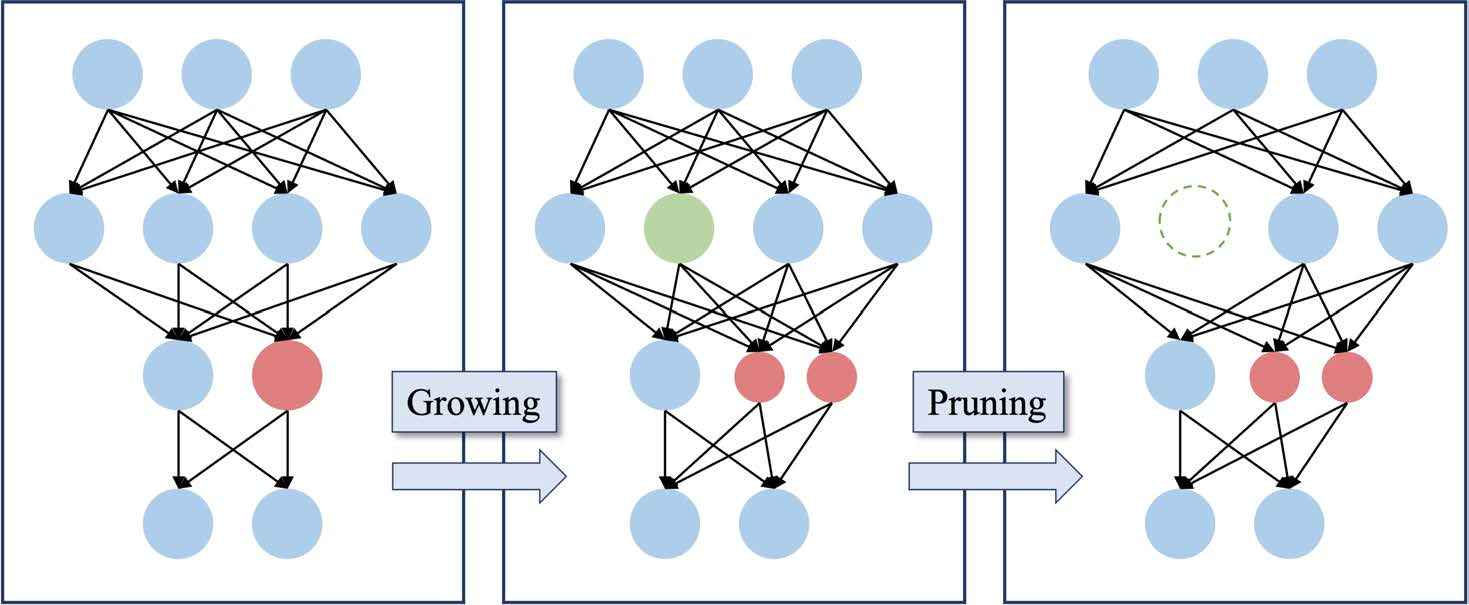}}
\caption{Two phases of growing and pruning.}
\label{fig3}
\vspace{-10pt}
\end{figure}

This section will introduce the growing and pruning method for model compression. 
Based on biological neural theories, neurons compete with each other. The neurons with more substantial competitiveness are suitable for survival and growth while the neurons with weaker competitiveness will be eliminated\cite{b21}\cite{b22}. Therefore, the model compression can be divided into two phases, the growing phase and the pruning phase, shown in Fig.~\ref{fig3}.

As shown in Fig.~\ref{fig3}, a growing phase is first performed to estimate how many neurons will be enough to support the network functions. It is worth explaining that although UEs are resource-constrained, the growing process is designed before our real implementation and does not have to be run online on UE devices. So the constrained resources will not be a problem. During the growing phase, the number of neurons is added one by one by splitting the neuron with the strongest competitiveness into two. If no improvement is observed in the network performance as the model size increases, the NN is supposed to be large enough to support the network functions. Then the algorithm steps into the pruning process and the neurons are pruned one by one by cutting off the neuron with the weakest competitiveness until there are no more neurons to be cut off.

The competitiveness of a neuron is defined according to the average percentage of zeros (PoZ) by measuring the percentage of zero activations of a neuron after the rectified linear activation unit (ReLU) mapping \cite{b23}, which is given as:
\begin{align}
PoZ_i^j = \frac{\sum_{z}^{Z}\sum_{l}^{L}f(o^{(i)}_{l,j}(z)==0)}{Z*L},
\label{eq13}
\end{align}
where $PoZ_i^j$ denotes the average PoZ of the $j^{th}$ neuron in the $i^{th}$ layer. $L$ and $Z$ denote the dimension of the neuron output and the number of validation samples. $o^{(i)}_{l,j}$ denotes the $l^{th}$ output of the $j^{th}$ neuron in the $i^{th}$ layer. $f(*)$ is a binary logic function, which can be given as:
\begin{align}
f(*)=\begin{cases}
&1,\ \ if\ true.
\\&0,\ \ if\ false.\label{eq14}
\end{cases}
\end{align}
Since the ReLu mapping will change all the negative inputs into zero, the average PoZ of Relu mapping can reflect the percentage of discarded and reserved neuron information. The neuron with the highest average PoZ is supposed to reserve the minor information and have the weakest competitiveness. In contrast, the neuron with the lowest average PoZ keeps the most information and have the most substantial competitiveness.

During the growing phase, when a neuron is split into two, the parameters of new neurons are initialized with the parameters of the old neuron. This can ensure that the split process will not influence the training process and decay the NN performance. The initialized weights of new neurons after splitting the $j^{th}$ neuron at the $i^{th}$ layer is given as:
 \begin{align}
 & w^{(i)'}_{l,j} = \Delta w^{(i)}_{l,j} \ \ \ \ \ \ \ \ \ \ \ \ \ \ \ \ \ \ \ \ \ \ \ \ \ \  \forall l \in N_{i-1},\label{eq15}
 \\& w^{(i)'}_{l,j+1} = (1-\Delta) w^{(i)}_{l,j} \ \ \ \ \ \ \ \ \ \ \ \ \ \ \ \ \ \forall l \in N_{i-1},\tag{17a}
 \\& w^{(i+1)'}_{j,k} = w^{(i+1)'}_{j+1,k} = w^{(i+1)}_{j,k} \ \ \ \ \ \ \ \ \ \ \forall k \in N_{i+1},\tag{17b}
 \\& \Delta \in (0,1)\nonumber
 \end{align}
where $w^{(i)}_{l,j}$ denotes the weight of the $l^{th}$ dimension of the $j^{th}$ neuron at the $i^{th}$ layer, and $w^{(i)'}_{l,j}$ denotes the updated weight value after neuron split. $N_i$ denotes the dimension of neurons at the $i^{th}$ layer and $w^{(i)'}_{l,j}$ denotes the updated weight after neuron split. $\Delta$ is a parameter for segmentation that can be chosen from zero to one. To make a proof that there is no decay in NN performance before and after the neuron split, we calculate the output of the $j^{th}$ neuron and the $(j+1)^{th}$ neuron at the $i^{th}$ layer with the initialized weight after the neuron split as follows:
\begin{align}
o_j^{(i)'} = &\sum^{N_{i-1}}_{l=1}w_{l,j}^{(i-1)'}o_l^{(i-1)} 
  = \Delta \sum^{N_{i-1}}_{l=1}w_{l,j}^{(i-1)}o_l^{(i-1)} 
  = \Delta o_j^{(i)} ,
 \end{align} 
 \begin{align}
o_{j+1}^{(i)'} = & \sum^{N_{i-1}}_{l=1}w_{l,j+1}^{(i-1)}o_l^{(i-1)}  = (1-\Delta) \sum^{N_{i-1}}_{l=1}w_{l,j}^{(i-1)}o_l^{(i-1)} \nonumber
 \\ = &(1-\Delta) o_j^{(i)}, 
 \end{align}
Since the format of ReLu is given as follows:
\begin{align}
ReLu(x) = x^+ = max(x,0),
\end{align}
we can make inferences from ReLu mapping as follows:
\begin{align}
\because\ &ReLu((1-\Delta) o_{j}^{(i)}) = (1-\Delta)ReLu( o_{j}^{(i)})
\\&ReLu(\Delta o_{j}^{(i)}) = \Delta ReLu( o_{j}^{(i)})\nonumber
\\\therefore\ &ReLu((1-\Delta) x_{j}^{(i)}) + ReLu(\Delta x_{j}^{(i)})
\\= &(1-\Delta)ReLu( o_{j}^{(i)}) + \Delta ReLu( o_{j}^{(i)}) \nonumber
\\= &ReLu(x_{j}^{(i)}),\nonumber
\end{align}

So, the output of any neurons at the $(i+1)^{th}$ layer can be given as:
\begin{align}
\forall k \in &N_{i+1},\nonumber
\\ o_k^{(i+1)'} = &\sum^{N_{i}}_{J=1}w_{J,k}^{(i+1)'}ReLu(o_{J}^{(i)'})
\\ = &\sum^{j-1}_{J=1}w_{J,k}^{(i+1)'}ReLu(o_{J}^{(i)'}) + w_{j,k}^{(i+1)'}ReLu(o_{j}^{(i)'}) \nonumber
\\+ &w_{j+1,k}^{(i+1)'}ReLu(o_{j+1}^{(i)'})
+ \sum^{N_{i}}_{J=j+2}w_{J,k}^{(i+1)'}ReLu(o_{J}^{(i)'})\nonumber
\\ = &\sum^{j-1}_{J=1}w_{J,k}^{(i+1)}ReLu(o_{J}^{(i)}) + w_{j,k}^{(i+1)}ReLu((1-\Delta) o_{j}^{(i)}) \nonumber
\\+ &w_{j,k}^{(i+1)}ReLu(\Delta o_{j}^{(i)})
+ \sum^{N_{i}}_{J=j+1}w_{J,k}^{(i+1)}ReLu(o_{J}^{(i)})\nonumber
\\ = &\sum^{j-1}_{J=1}w_{J,k}^{(i+1)}ReLu(o_{J}^{(i)}) +
w_{j,k}^{(i+1)}ReLu(o_{j}^{(i)})\nonumber
\\+ &\sum^{N_{i}}_{J=j+1}w_{J,k}^{(i+1)}ReLu(o_{J}^{(i)})\nonumber
\\ = &o_k^{(i+1)},\nonumber
\end{align}
where it can be observed that the output of any neurons at the $(i+1)^{th}$ layer after neuron split is the same as the output before the neuron split. It proves that there is no performance decaying after the neuron split. The model compression procedure is summarized at Algorithm \ref{alg:modelC}. 

\begin{algorithm}[t]
\caption{Model compression for hidden neuron number determination}\label{alg:modelC}
\begin{algorithmic}[1]
\Initialize{$N_1,N_2 \gets 2$, $\theta_m^{local}$, $S_m$ and $A_m$ of each UE, $\overline{P_{T-1}} \gets -\infty$,$N_{split} \gets 0$,$\overline{P_{T}} \gets 0$}
\While{$mod(TTI, SplitInterval) != 0$}
\For{each UE}
    \State Do local training via Algorithm \ref{alg:localT}
    \State $\overline{P_{T}} \gets \overline{P_{T}} + R_0$
\EndFor
\State $TTI \gets TTI + 1$
\EndWhile

\State $TTI \gets TTI + 1$
\If{$N_{split} \leq N_{required}$}
    \If{$P_{T} < P_{T-1}$}
    \State $N_{split} \gets N_{split} + 1$
    \Else
    \State $N_{split} \gets 0$
    \EndIf
\EndIf

\State Calculate $PoZ_i^j$ in $SplitInterval$ via Equation (\ref{eq13})
\If{$N_{split} \leq N_{required}$}
\State $j_i = \underset{j}{\max}PoZ_i^j\ \ \forall{i} = 1,2$
\State $N_1\gets N_1+1$
\State $N_2\gets N_2+2$
\State Split the $j^{th}$ neuron at the $\forall{i} = 1,2$ layer and initialize $w^{(i)'}_{l,j}$, $w^{(i)'}_{l,j+1}$ and $w^{(i+1)'}_{l,j}$ via Equation (\ref{eq15})
\State $P_{T-1} = P_{T}$
\State \Goto\ Line 2
\Else
\State $i,j = \underset{i,j}{\min}PoZ_i^j$
\If{$i==1$ and $N_1 > 2$}
\State $N_1\gets N_1-1$
\Else
\State $N_2\gets N_2-1$
\EndIf
\State Cut off the $j^{th}$ neuron at the $i^{th}$ layer and record $N_1$, $N_2$ and $P_{T}$
\If{$N_1 >2$ or $N_2 > 2$}
\State \Goto\ Line 2
\EndIf
\EndIf

\end{algorithmic}
\end{algorithm}

\vspace{-10pt}
\subsection{Q-value-based Knowledge Transfer}

\begin{figure}[t]
\centerline{\includegraphics[width=3.4in]{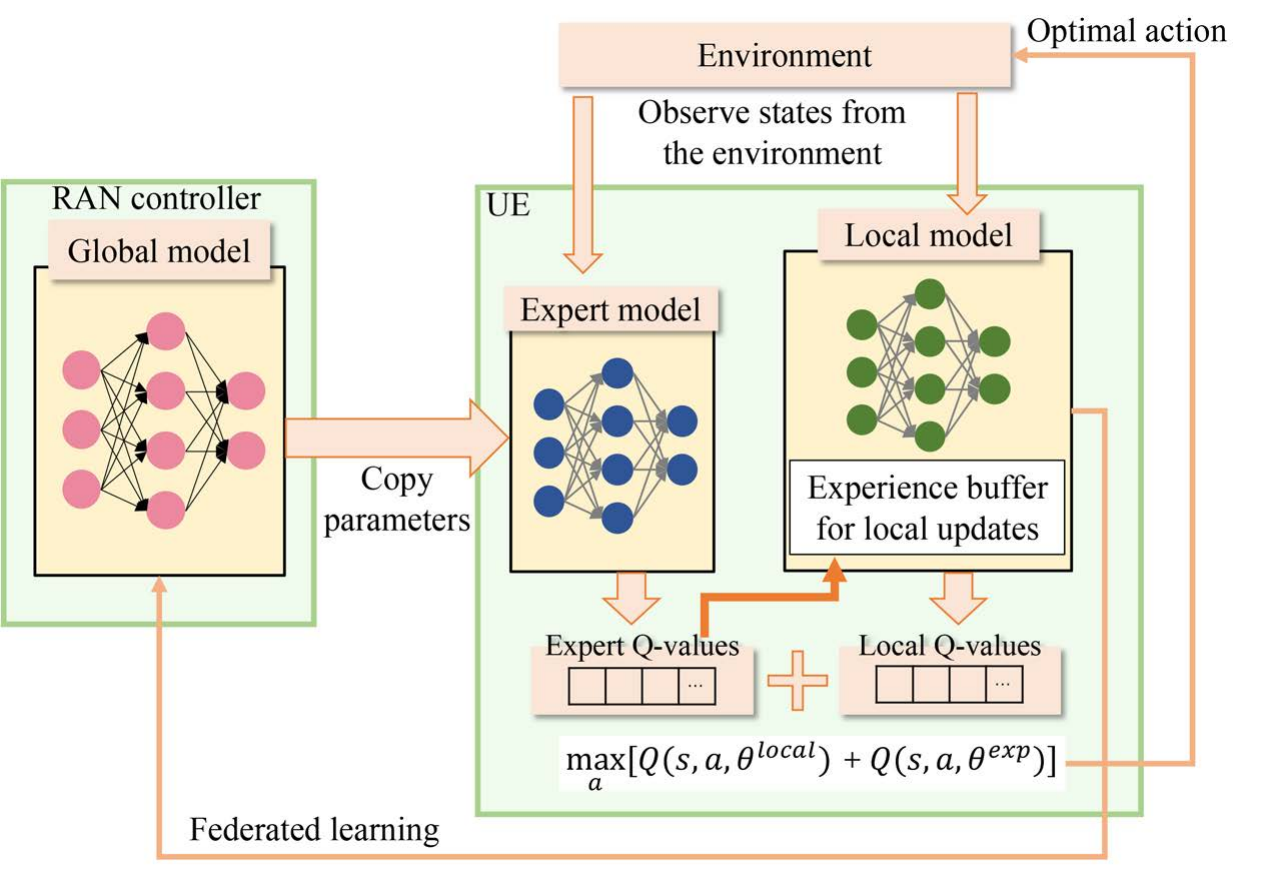}}
\caption{Knowledge transfer process.}
\label{fig5}
\vspace{-10pt}
\end{figure}
This section introduces the transfer learning technique. The framework of knowledge transfer is shown in Fig.~\ref{fig5}. We define two models for each UE, an expert model and a local model. During the TS process, both the expert and local models of a UE observe the state from the environment and generate Q-vales for different actions. The expert model is initialized by the global model in the RAN controller through parameter replication. Although expert models share the same network parameters with the global model, they are different models because they are deployed at different parts of the network and play different roles. The expert model is deployed inside each UE. It takes the previously learned knowledge of the global model as expert knowledge and provides expert Q-values based on the global experience during local training process in federated learning. The local model provides local Q-values based on the local experience. The TS actions are chosen by maximizing the sum of expert and local Q-values. This process can be formulated as:
\begin{align}
a = \underset{a}{argmax}(Q(s_{m,t},A;\theta_{m,t}^{local})+Q(s_{m,t},A;\theta_{m,t}^{exp})),
\label{eq27}
\end{align}
where $\theta^{local}$ denotes local model parameters and $\theta^{exp}$ denotes expert model parameters. In this way, the global experience from the expert model and the local experience from the local model are combined for decision-making.

When a new UE arrives at the system, the global model parameters are directly copied into the local and expert models for initialization. Next, during each federated interval, the actions are chosen as a combined effort of both the local and expert models. We use separated local models and expert models since the local model is trained and updated during local training, while the global model remains unchanged. When federated learning is performed, the global model is updated by federating all the local models. The updated global model is used to update the expert models at each UE side. Therefore, the global experience and local experience can be simultaneously learned but separately kept at each UE. Another advantage of using the proposed knowledge transfer scheme is that it can reduce the running time by performing global model aggregation and local training simultaneously. Conventionally, the local model update is based on feedback from the global model, so local training and model aggregation should be performed sequentially. But with the introduction of the expert model, feedback is given to the expert model instead of the local model. That means UEs do not need to wait for feedback from the global model and model aggregation can be run in parallel with local training. In this way, the running time of model aggregation is saved.

Since the global model is an aggregation of local models and expert models are transferred from the global model, the local models and expert models are supposed to converge to the same value. Therefore, we redefine the loss function of local training as:
\begin{align}
Loss_{m} = &\underset{\Psi_m}{\sum}(y_m + Q(s_{m,t},a_{m,t};\theta_{m,t}^{exp})-2R_{m,t} \nonumber
\\&-\gamma\underset{a}{\max}({Q(s_{m,t+1},a};\theta_{m,t}^{local})+Q(s_{m,t+1},a;\theta_{m,t}^{exp}))^2.\label{eq26}
\end{align}

Since the expert Q model is not updated during the local training, $Q(s_{m,t+1},a;\theta_{m,t}^{exp})$ and $Q(s_m^t,a_m^t;\theta_{m,t}^{exp})$ can be seen as constants. So the update of local models at each UE can be formulated as:
\begin{align}
\theta_{m,t+1} = &\theta_{m,t}+\alpha[2R_{m,t}-Q(s_{m,t},a_{m,t};\theta_{m,t}^{exp})\nonumber
\\&-\gamma\underset{a}{\max}({Q(s_{m,t+1},a};\theta_{m,t}^{local})+Q(s_{m,t+1},a;\theta_{m,t}^{exp}))\nonumber
\\&-Q(s_{m,t},a_{m,t};\theta_{m,t}^{local})]\nabla{Q(s_{m,t},a_{m,t}};\theta_{m,t}^{local}).\label{eq28}
\end{align}

With this definition, the sum of expert Q-values and local Q-values is supposed to converge to twice the long-term accumulated reward. On the other hand, expert Q-values and local Q-values will converge to a similar value according to the federated learning process. So the final output is expected to converge to the long-term accumulated reward of given actions and states. In this way, expert models can provide a lead for UEs according to the previous experience of other UEs and UEs can make personalized modifications based on the suggestions of experts. It is worth pointing out that the expert models do not need to be trained on the UE side. It is only used for model inference during the local training, which only requires minimum computation.

As a combination of the aforementioned grouped attention-weighted federated learning and transfer learning, the complete version of the proposed federated reinforcement learning-based UE-centric TS is summarized at Algorithm \ref{alg:ftl}.

\begin{algorithm}[t]
\caption{KT-FLUC-based TS}\label{alg:ftl}
\begin{algorithmic}[1]
\Initialize{$\theta_{local},\theta_{local}^{exp}$ of each UE, $S^{t-1}$ and $a^t$}
 \While {$TTI < T^{total}$}
    \State $TTI \gets TTI + 1$
    \For{each UE}
        \State Observe local state.
        \State $S \gets \{\Gamma_{k},L^{queue}_k,delay_n,k_n|k\in K\}$.
        \State Get immediate reward $R$.
        \State Record experience $\{S_{m,t+1},S_{m,t},a_{m,t},R_{m,t}\}$.
        \State Sample a minibatch from experience buffer
        \State Update local Q network $\theta$ with 
        \State $\nabla(L_{exp} (\theta_{local},\theta_{exp}))$ via equation (\ref{eq26}).
        \State Choose actions via equation (\ref{eq27}).
        \State Execute action $A$.
    \EndFor
    \If{$mod(TTI,FedInterval)==0$}
        \For{each UE} 
            \State Submit $\theta_{local}$ to RAN controller.
            \State $\overline{R}_{m} \gets \sum_{TTI \in FedInterval}R_{TTI}/FedInterval$
            \State $|{MB}_{m}| \gets length\{buffer\}/T^{total}$
            \State ${\Phi}_{m} \sum_{TTI \in FedInterval}h_m/FedInterval$
            \State $w_m \gets softmax(\overline{R}_{m},|{MB}_{m}|,{\Phi}_{m})$
        \EndFor
        \State $\theta_{global} \gets (1-\eta_1)\theta_{local} + \eta_1 w_m \sum_{m \in M}\theta_{m}/|M|$ 
        \For{each UE} 
            \State $\theta_{local}^{exp} \gets \theta_{global}$
        \EndFor
    \EndIf
    \If{A new UE enters the network.}
        \State $\theta_{local} \gets \theta_{global}$
        \State $\theta_{local}^{exp} \gets \theta_{global}$
    \EndIf
\EndWhile
\end{algorithmic}
\end{algorithm}

\vspace{-10pt}
\subsection{Baselines}
In this paper, we proposed four baseline algorithms. Firstly, we compared our proposed KT-FLUC algorithm with a centralized cell-centric DRL-based TS algorithm (CL), which is implemented based on \cite{b9}. In this algorithm, we only use a single agent at the RAN controller, collecting observations from distributed UEs and making a centralized decision. The action is a joint TS decision of all the UEs, which will be broken into independent TS actions for each UE, and the model will be trained based on the average reward of all the UEs. The critical steps of the CL algorithm are summarized at Algorithm \ref{alg:cl}.

\begin{algorithm}[t]
\caption{Federated DRL based TS with direct initialization}\label{alg:fli}
\begin{algorithmic}[1]
\Initialize{$\theta_{local}$ of each UE, $S_{t-1}$ and $a$}
 \While {$TTI < T^{total}$}
    \State $TTI \gets TTI + 1$
    \For{each UE}
        \State Do local training via Algorithm \ref{alg:localT}
    \EndFor
    \If{$mod(TTI,FedInterval)==0$}
        \For{each UE} 
            \State Submit $\theta_{local}$ to RAN controller.
            \State $\overline{R}_{m} \gets \sum_{TTI \in FedInterval}R_{TTI}/FedInterval$
            \State $|{MB}_{m}| \gets length\{buffer\}/T^{total}$
            \State ${\Phi}_{m} \sum_{TTI \in FedInterval}h_m/FedInterval$
            \State $w_m \gets softmax(\overline{R}_{m},|{MB}_{m}|,{\Phi}_{m})$
        \EndFor
        \State $\theta_{global} \gets (1-\eta_1)\theta_{local} + \eta_1 w_m \sum_{m \in M}\theta_{n}/|M|$ 
        \For{each UE} 
            \State $\theta_{local} \gets (1-\eta_2)\theta_{local} + \eta_2\theta_{global}$
        \EndFor
    \EndIf
    \If{A new UE enters the network.}
        \State $\theta_{local} \gets \theta_{global}$
    \EndIf
\EndWhile
\end{algorithmic}
\end{algorithm}

\begin{algorithm}[t]
\caption{Cell-centric DRL-based TS}\label{alg:cl}
\begin{algorithmic}[1]
\Initialize{$\theta_{local}$ of each UE, $S_p$ and $a$}
 \While {$TTI < T^{total}$}
    \State $TTI \gets TTI + 1$
    \For{each UE}
    \State Observe local state.
    \State $S^{t}_m \gets \{\Gamma_{n},L^{queue}_n,d_m,\beta_{m,n}|n\in N\}$.
    \State Get immediate reward $R^t_mR$.
    \EndFor
    \State Collect local states and get cell-based state $S^{t}_c$.
    \State Add rewards from all the UEs and get cell-based reward $R^{t}_c$.
    \State Record experience $\{S^{t-1}_c,S^{t}_c,a^t_c,R^t_c\}$.
    \State Sample a minibatch from experience buffer
    \State Update Q network $\theta$ 
    \State Choose actions by $a^t_c \gets \underset{a^t_c}{argmax}Q(S,a^t_c;\theta)$.
    \State Separate the cell-based action $a^t_c$ into actions for each cell $a^t_m$.
    \State Execute action $a^t_m$.
\EndWhile
\end{algorithmic}
\end{algorithm}

Then, we compared our proposed KT-FLUC algorithm with three other UE-centric algorithms. The first one is a distributed independent DRL-based TS algorithm (DIL), which is implemented based on the idea of \cite{b23-1}. In this baseline, each UE acts as an independent agent, making decisions without information exchanging or experience sharing. The activities of each UE in DIL are summarized at Algorithm \ref{alg:localT}. 

The third baseline is a federated DRL-based TS algorithm (FL). The federated learning algorithm used in this baseline is borrowing ideas from \cite{b23-2}. Compared with DIL, the federated learning steps are added, while compared with the proposed algorithm, the knowledge transfer part is removed. The FL algorithm is summarized at Algorithm \ref{alg:fl}. The last baseline is a federated DRL-based TS algorithm with direct initialization (FLI). Direct initialization means that the global model parameters are directly copied to the local model every time a new UE enters the system. The FLI algorithm is summarized at Algorithm \ref{alg:fli}.
\vspace{-10pt}
\subsection{Computational Complexity Analysis}

In this section, we analyze the computational complexity of the proposed KT-FLUC framework. For each UE, the complexity of the DQN model deployed on the UE is dominated by the training and updating of NN, consisting of updating the weights of neurons and the bias. So the computational complexity is proportional to the number of NN layers, the dimension of neurons and the number of neurons on each layer, which can be given as $O(\sum^{I}_{i=1}N_{i-1}N_i)$, where $N_i$ denotes the NN size at the $i^{th}$ layer. In addition, the knowledge transfer process also contributes to the complexity. The computation complexity of summing local and expert Q-values is proportional to the size of action space, which is given as $O(|A|)$. $A$ denotes the action set of the UE.

For the RAN controller, the complexity is dominated by the federated learning process. The computation complexity of the attention weight calculation can be formulated as $O(M)$, where $M$ denotes the number of UEs. The complexity of global update is given as $O(M\sum^{I}_{i=1}N_{i-1}N_i)$. In summary, these analyses show that the computation complexity of our KT-FLUC framework is proportional to the NN size and the UE number. 

In summary, these analyses show that the computational complexity of local training is proportional to the number of neurons. In practical implementation, the UE usually has limited memory and computation capacity, while it will take them significant processing of network parameters to perform applications deployed on to them\cite{b26}. The growth of NN will have a critical impact on energy, power, and timing. So it is necessary to limit the NN size. In this paper, we have kept the number of neurons as low as possible through neural network compression to save the computation cost of local training. The complexity of global updates is proportional to the number of UEs and the number of neurons. Although the complexity will increase when the number of UE increases, the global server is usually assumed to be powerful enough to deal with the high computational complexity. On the other hand, the global model aggregation is a non-real-time process, so it will not be computationally demanding. Therefore, the system has good scalability and can still work well when there are a large number of UEs.

\begin{table}[t]
\caption{Simulation settings.}
\label{table1}
\vspace{-5pt}
\renewcommand{\arraystretch}{1.2}
\centering
\begin{adjustbox}{width=\columnwidth,center}
\begin{tabular}{|l|l|}
\hline
\textbf{Network environment settings}                                                   & \textbf{UE traffic settings}                \\ \hline
Bandwidth: 20 MHz                                                                       & \textbf{GBR traffic}                        \\
Subcarriers of each RB: 12                                                              & QCI: 2                                      \\
Tx/Rx antenna gain: 15 dB                                                               & Proportion: 40\%                            \\
Number of resource blocks: 100                                                          & Data length: 50 KB                          \\
\begin{tabular}[c]{@{}l@{}}Propagation model: \\ 128.1 + 37.6log(Distance)\end{tabular} & Packet size: 1600 Bytes                     \\
Log-Normal shadowing: 8 dB                                                              & Packet arrival mode: Constant bit rate data \\
Number of UEs: 25/35/45/55/65                                                           & Inter-arrival time between packets: 3 ms    \\ \cline{2-2} 
Number of BSs: 5.                                                                       & \textbf{Non-GBR traffic}                    \\ \hline
\textbf{MBS settings}                                                                   & QCI: 6                                      \\ \cline{1-1}
Radio access technology: LTE                                                            & Proportion: 60\%                            \\
Carrier frequency: 0.8 GHz                                                              & Data length: 250 KB                         \\
Transmission power: 40 W                                                                & Packet size: 3200 Bytes                     \\
Number of MBS: 1                                                                        & Packet arrival mode: Constant bit rate data \\
Bandwidth: 10 MHz                                                                       & Inter-arrival time between packets: 3 ms    \\ \cline{2-2} 
Channel model: 3GPP urban macro                                                         & \textbf{Deep learning settings}             \\ \hline
\textbf{SBS settings}                                                                   & Initial learning Rate: 0.001                \\ \cline{1-1}
Radio access technology: 5G NR                                                          & Discount factor: 0.5                        \\
Carrier frequency: 3.5 GHz                                                              & Epsilon value: 0.05                         \\
Transmission power: 20 W                                                                & Experience pool: 200                        \\
Number of MBS: 4                                                                        & Batch size: 64                              \\
Bandwidth: 20 MHz                                                                       & Hidden layer: 2                             \\
Channel model: 3GPP urban micro                                                         & Federated interval: 30 TTIs                 \\ \hline
\end{tabular}
\end{adjustbox}
\end{table}

\vspace{-10pt}
\section{Simulation Settings and Results}\label{s5}

\subsection{Simulation Settings} 
We consider a scenario with $N=5$ BSs, an LTE MBS in the center and 4 5G SBSs in the surrounding. We consider a dynamic system and UEs will arrive and depart during the simulation. %We suppose the traffic type of a single UE is fixed and the traffic load of UEs with the same traffic type is similar. So the traffic load of the system is controlled by the average number of UE in the system. 
40\% UEs have GBR service and 60\% UEs have non-GBR service. The simulation is repeated five times using MATLAB, and the average outcomes are presented along with a 95\% confidence interval. The $B_{max}$ and $D_{max}$ we use during the simulation are 10 ms and 10 Mbps. The update speed of the global model $\eta_1$ is 0.5. The update speed of the local model $\eta_2$ is 1. The split interval we use during model compression is 150 TTIs. The $N_{required}$ is set as 5.
During the simulation, we change the average number of UEs from $M=25$ to $M=65$ to see how the proposed algorithm performs under different UE numbers and traffic loads. Depending on the dimensions of the state space and action space, the number of hidden neurons is initially set to 32 and 64 for most UEs. Then we perform growth and pruning for one single UE and start with two hidden neurons in each layer.
Other simulation details about network environment settings and deep learning settings are shown in Table \ref{table1}.

\begin{figure}[t]
\centerline{\includegraphics[width=8cm,height=5.8cm]{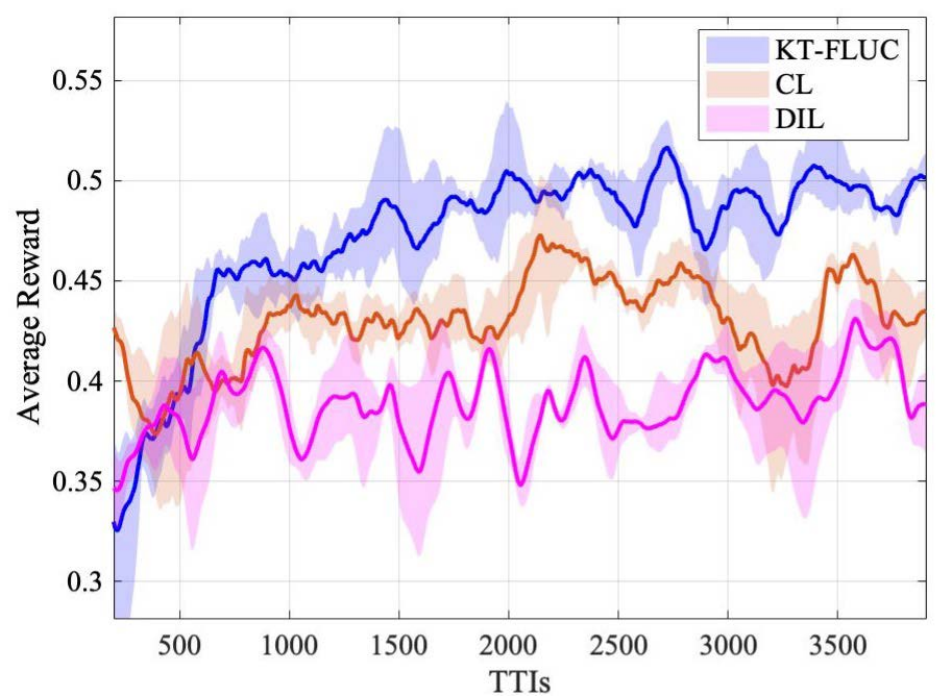}}
\vspace{-5pt}
\caption{The convergence performance of system reward.
%\red{can you reduce the gap from x-axis to the figure title? Is it because you have some white space in the original figure? Btw, enlarge the fonts as Fig.7.}
}
\label{fig10}
\vspace{-10pt}
\end{figure}

\begin{figure}[t]
    \centering
    \subfigure[The average GBR traffic delay with different UE numbers.]{
	\includegraphics[width=8cm,height=5.8cm]{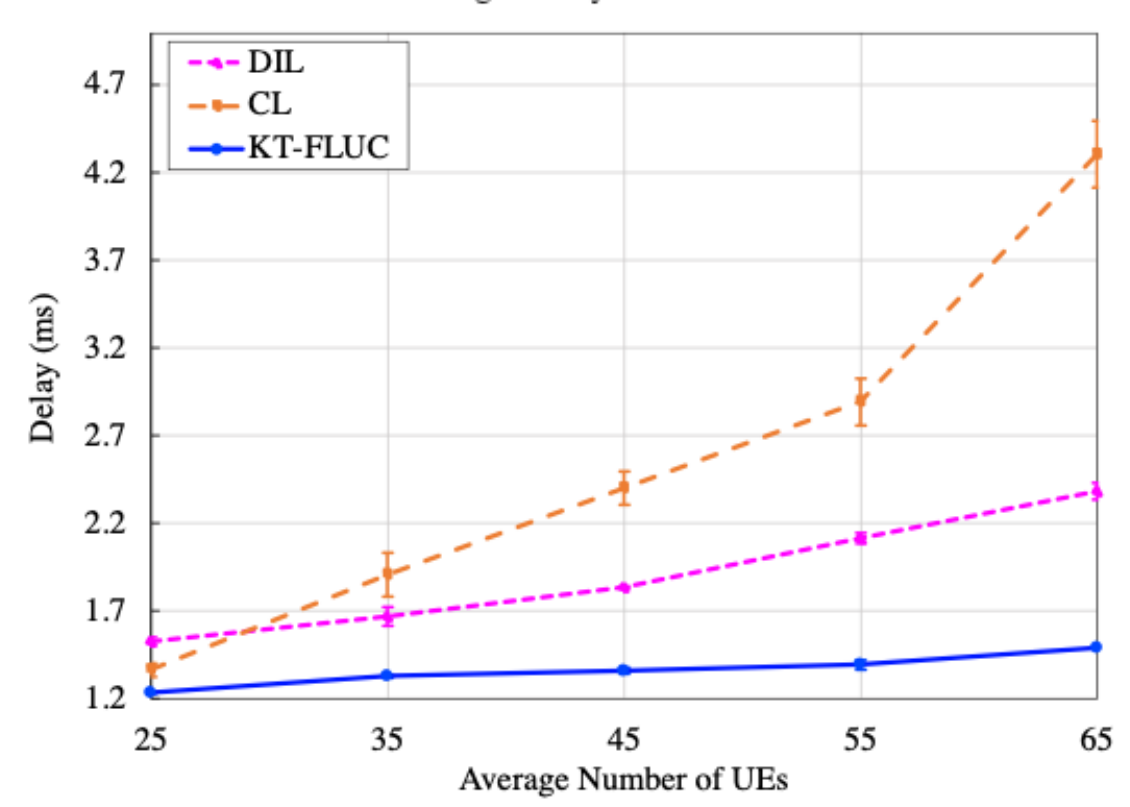}
    }
    \centering
    \subfigure[The average non-GBR traffic throughput with different UE numbers.]{
    	\includegraphics[width=8cm,height=5.8cm]{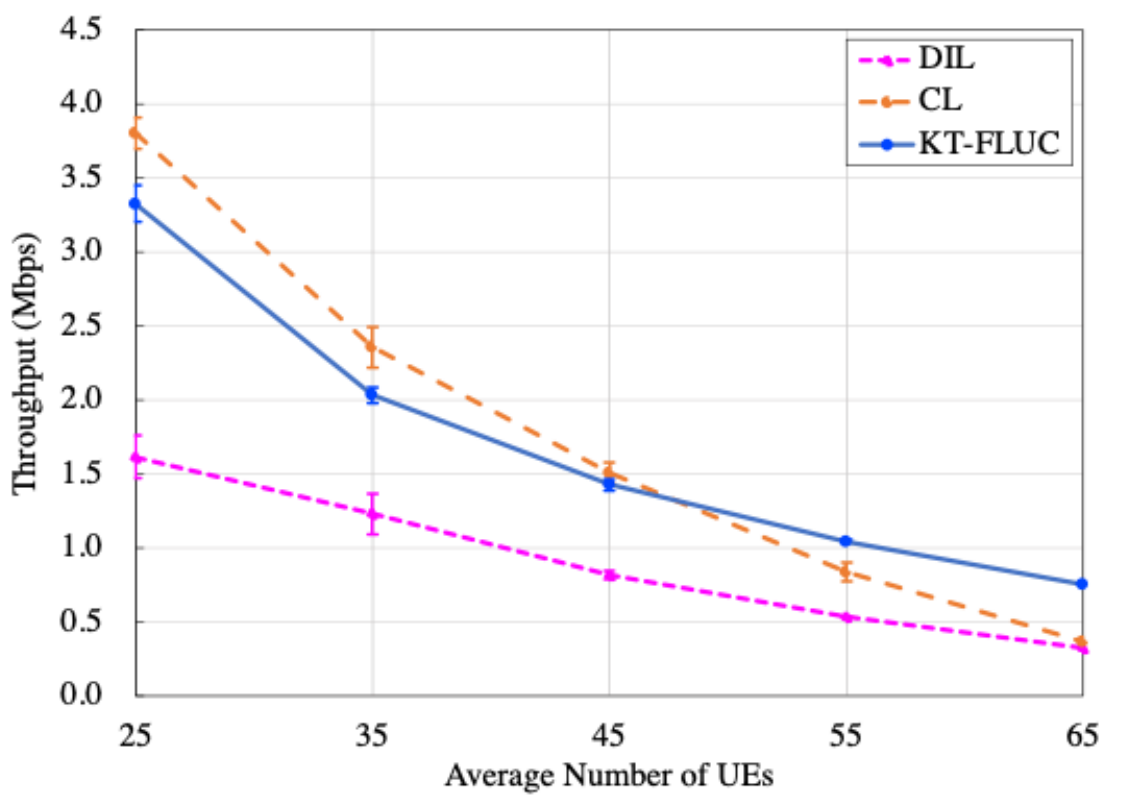}
    }
    \caption{The network performance of KT-FLUC, CL and DIL. (a) The average GBR traffic delay. (b) The average non-GBR traffic throughput.}
    \label{fig11}
\vspace{-10pt}
\end{figure}

\vspace{-10pt}
\subsection{Performance Analysis between UE-centric TS and Cell-centric TS}
% In this section, we compare the simulation results of \red{cell-centric baseline} CL and the proposed KT-FLUC algorithm to analyze the advantages of using a UE-centric TS structure instead of a cell-centric TS structure. 
This section will compare the proposed KT-FLUC with cell-centric baseline CL and DIL. 

Fig. \ref{fig10} shows the convergence curves of DIL, CL and proposed KT-FLUC algorithms when the number of UE is 45. It can be observed that the CL algorithm can achieve a higher system reward than the DIL algorithm, but a lower reward than our proposed KT-FLUC algorithm. This demonstrates that the cell-based algorithm has an advantage in system performance compared with simple UE-centric algorithms because the cell-centric algorithms hold more comprehensive information and can coordinate between UEs.

%By using some additional techniques to organize UEs, it is possible to make UE-centric TS algorithms perform better than cell-centric TS algorithms. This is because training smaller models on each UE is easier than training a sizeable centralized model. In the cell-centric system, UEs may not respond promptly to environmental changes considering the inability to extract adequate information from the massive volume of information especially in a highly dynamic environment. 

In Fig. \ref{fig11}, the average delay of GBR traffic and throughput of non-GBR traffic under different numbers of UEs are shown. As the number of UE increases, the competition for resources among UEs becomes more intense, so the average delay will increase, and the average throughput will decrease. It can be observed that the CL algorithm performs better when there are fewer UEs. In contrast, UE-centric algorithms are influenced less by the increase in  the number of UEs. This is because fewer UEs are easier to coordinate, and there is less information for the centralized model to deal with. When the average UE number is 25, the CL algorithm can achieve 14\% higher throughput compared with the proposed KT-FLUC algorithm. However when the average UE number is 65, the proposed KT-FLUC algorithm can achieve 52\% higher throughput than the CL algorithm. It can be observed that the delay of the CL algorithm is the highest among the three algorithms. This is because the cell-based algorithm tends to focus more on the overall performance of a cell rather than on the individual performance of UEs. Since the proportion of non-GBR traffic is much higher than GBR traffic, the CL algorithm may make decisions that are more suitable for non-GBR traffic at the expense of the benefits of GBR traffic to guarantee the overall reward. In comparison, in UE-centric algorithms, each UE fights for its own rewards, so it will choose the most appropriate actions for itself and makes the delay lower.
When the average UE number is 65, the proposed KT-FLUC algorithm can achieve a 65\% lower delay than the CL algorithm.

% It also can be observed that the throughput of CL is relatively high, close to the throughput of FTL, while the delay of CL is also very high, higher than the delay of DIL. This is because the cell-based algorithm tends to focus more on the overall performance of a cell other than on the individual performance of UEs. Since the proportion of non-GBR traffic is much higher than GBR traffic, the CL algorithm may make decisions that are more suitable for non-GBR traffic at the expense of the benefits of GBR traffic to guarantee the overall reward. In comparison, in UE-centric algorithms, each UE is fighting for their own rewards so they will choose the most appropriate actions for themselves other than making concessions for overall rewards. So the dthe elay of UE-centric algorithms is lower.

\begin{figure}[t]
    \centering
    \subfigure[The average traffic blocking of UEs of cell-centric TS and UE-centric TS.]{
        \includegraphics[width=8cm,height=3cm]{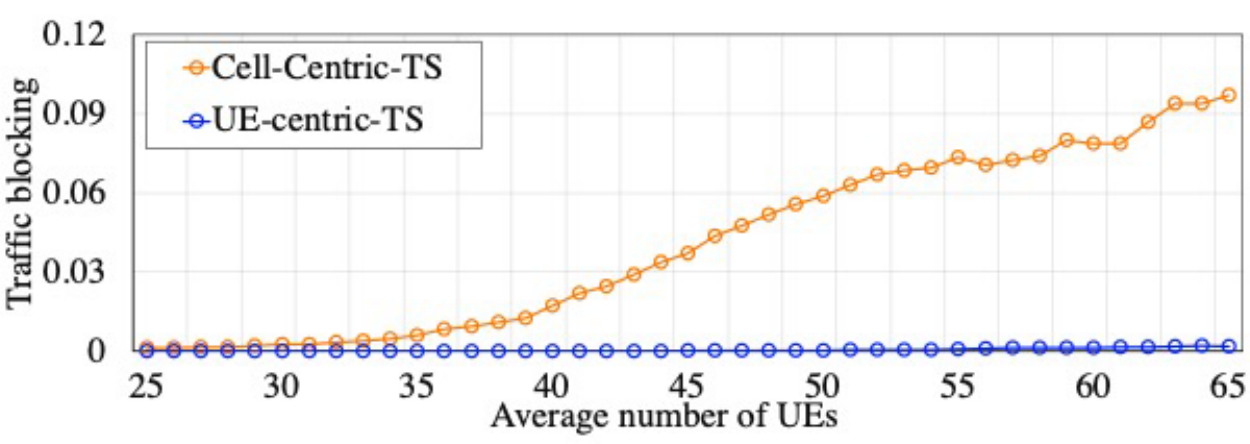}
    }
    \subfigure[The average QoS violation rate of UEs of cell-centric TS and UE-centric TS.]{
	\includegraphics[width=8cm,height=3cm]{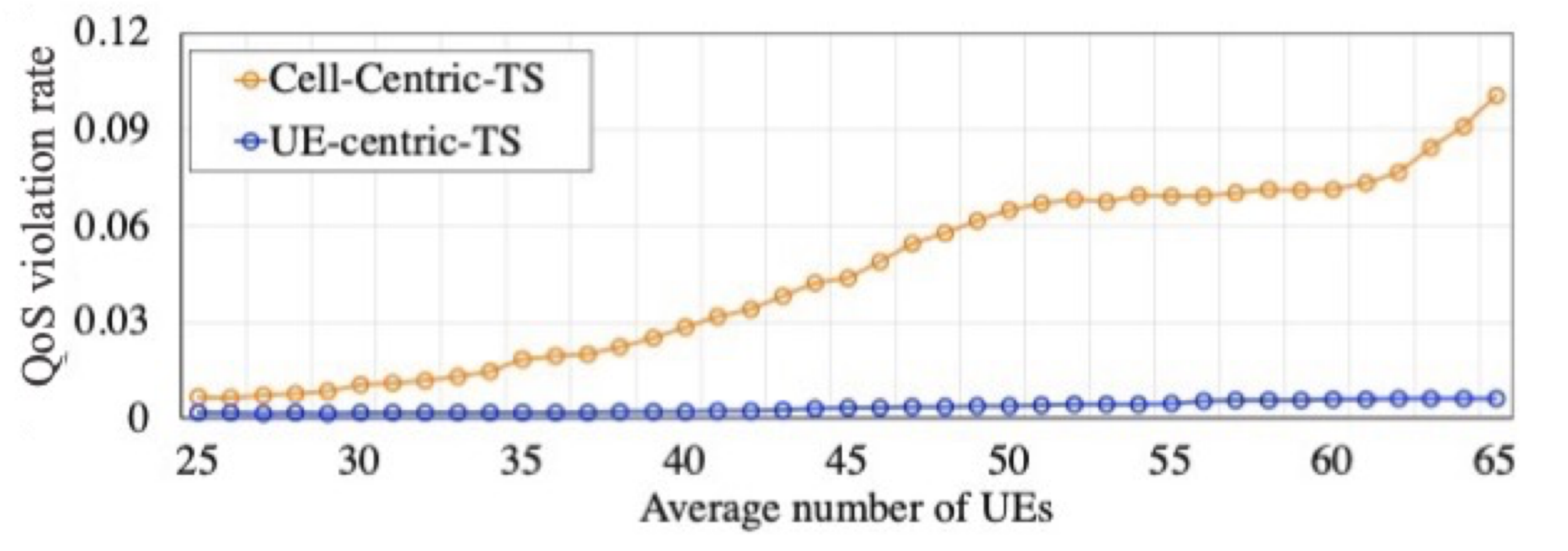}
    }
    \centering
    \subfigure[The average transmission cycle of UEs of cell-centric TS and UE-centric TS.]{
    	\includegraphics[width=8cm,height=3cm]{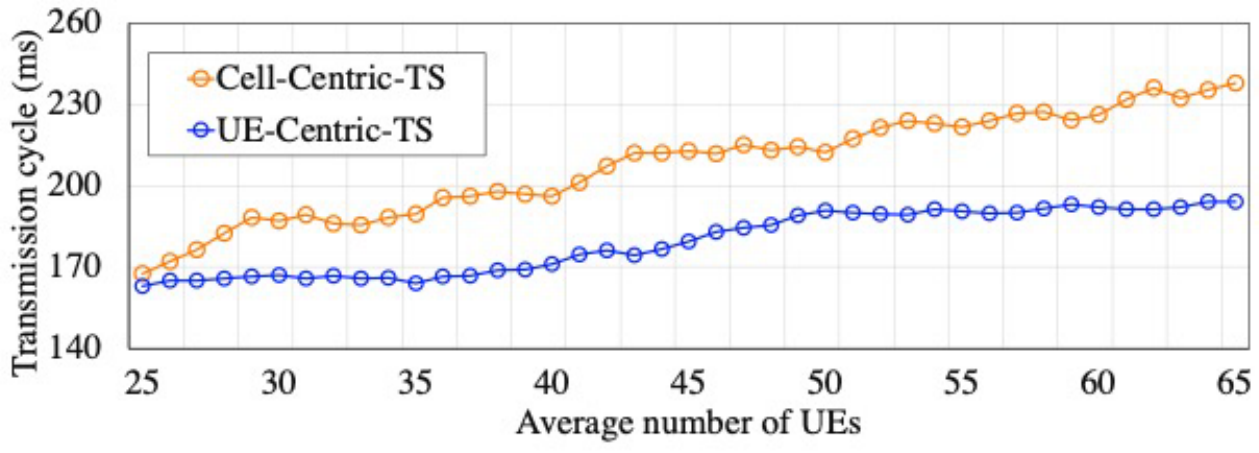}
    }
    \caption{The indicators of UE satisfaction as the number of UEs increases. (a) The average transmission cycle. (b) The average QoS violation rate. (c) The average transmission cycle.}
    \label{fig12}
    \vspace{-10pt}
\end{figure}

To further compare the cell-centric TS and UE-centric TS, we define three indicators to measure the UE's satisfaction with the TS service, traffic blocking, QoS violation rate and the transmission cycle. Since GBR traffic has a higher priority than non-GBR traffic, non-GBR traffic may be blocked when two kinds of traffic request transmission simultaneously. The traffic blocking represents the probability of GBR traffic being blocked. QoS violation rate represents the probability that the delay of GBR traffic exceeds the QoS requirements. And the last indicator, the transmission cycle, represents the average time for UEs to finish the transmission of a given file.

It can be observed from Fig. \ref{fig12} that the UE-centric TS can achieve a lower traffic blocking and lower QoS violation rate compared with cell-centric TS algorithms. This verifies that the UE-centric TS algorithm can respond more promptly to environmental changes. UE-based performance can give a more intuitive illustration of the UE's satisfaction than cell-based performance. The transmission cycle of UE-centric TS is also smaller than cell-centric TS, which means the traffic is transmitted faster in a UE-centric framework with more rational TS strategies.

\vspace{-10pt}
\subsection{Performance Analysis between Four UE-centric Frameworks}

% \begin{figure*}[h]
%     \centering
%     \subfigure[The average system reward with different UE numbers.]{
%         \includegraphics[width=2.25in]{UE-1.png}
%     }
%     \subfigure[The average GBR traffic delay with different UE numbers.]{
% 	\includegraphics[width=2.25in]{UE-2.png}
%     }
%     \centering
%     \subfigure[The average non-GBR traffic throughput with different UE numbers.]{
%     	\includegraphics[width=2.25in]{UE-3.png}
%     }
%     \caption{The network performance under different number of UEs of FTL, FLI, FL and DIL algorithms. (a) The average GBR traffic delay. (b) The average non-GBR traffic throughput. (c) The system reward.}
%     \label{fig9}
% \end{figure*}
In this section, we compare the simulation results from DIL, FL, FLI and proposed KT-FLUC algorithms to analyze the contribution of transfer learning and federated learning techniques and to verify that the combination of two algorithms can create synergy.

\begin{figure}[t]
\centerline{\includegraphics[width=8cm,height=6cm]{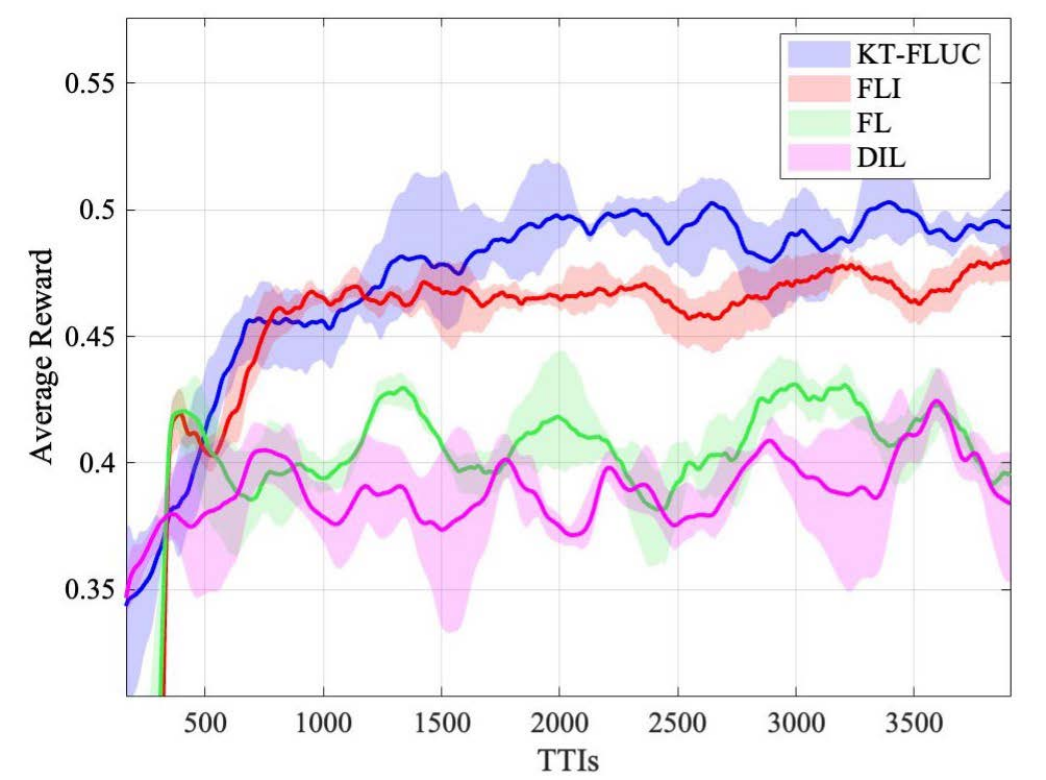}}
\caption{The performance of system reward.}
\label{fig8}
\vspace{-10pt}
\end{figure}

\begin{figure}[t]
    \centering
    \subfigure[The average GBR traffic delay with different UE numbers.]{
	\includegraphics[width=8cm,height=6cm]{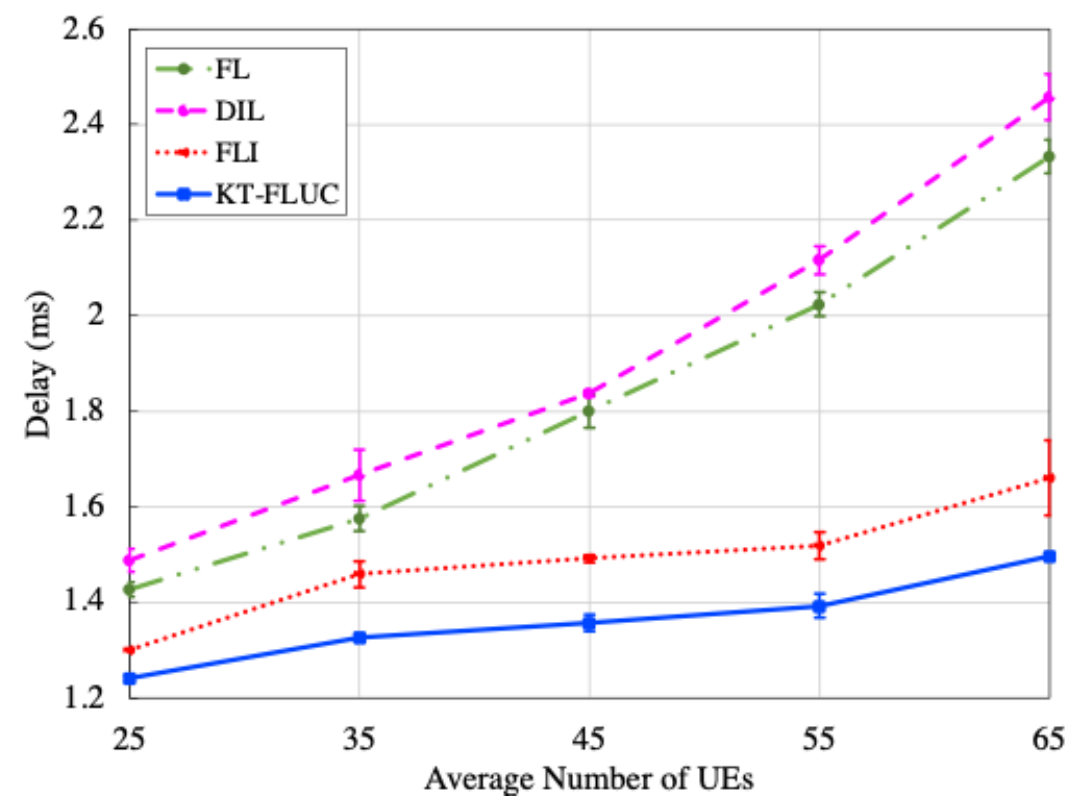}
    }
    \centering
    \subfigure[The average non-GBR traffic throughput with different UE numbers.]{
    	\includegraphics[width=8cm,height=6cm]{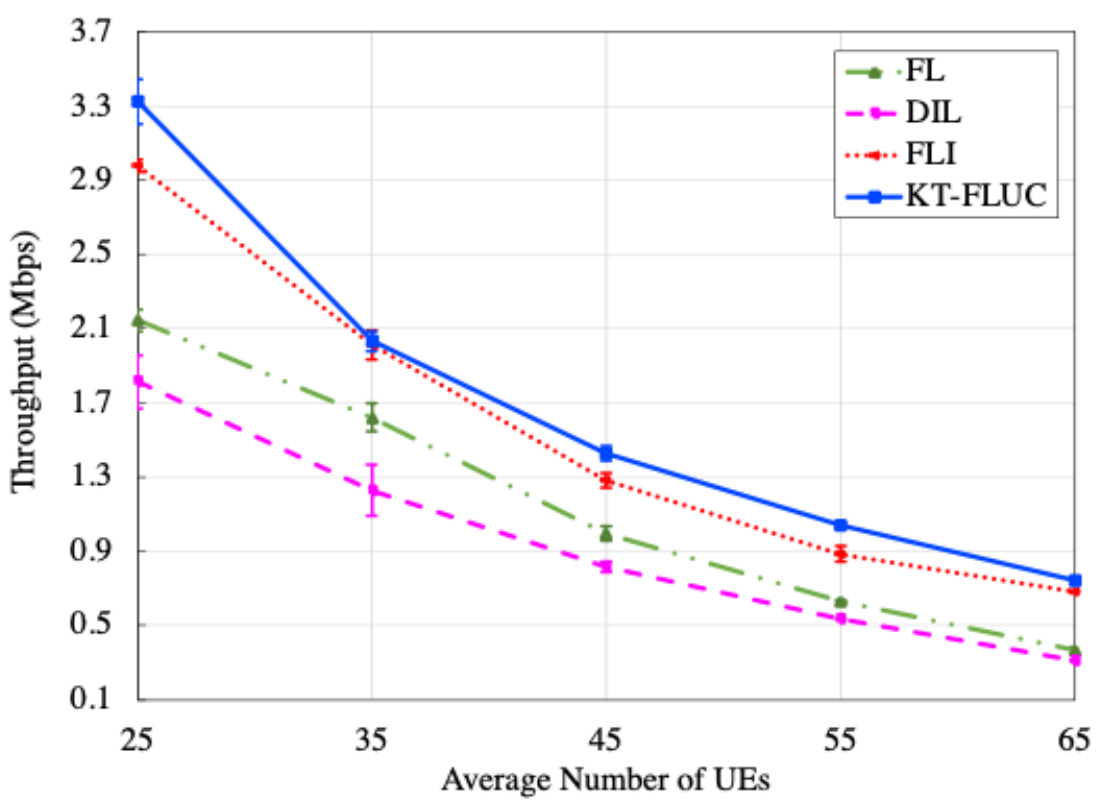}
    }
    \caption{The network performance of KT-FLUC, FLI, FL and DIL. (a) The average GBR traffic delay. (b) The average non-GBR traffic throughput.}
    \label{fig9}
    \vspace{-15pt}
\end{figure}

Fig. \ref{fig8} shows the convergence curves of DIL, FL, FLI and proposed KT-FLUC algorithms when the number of UE is 45. It can be observed that the DIL algorithm achieves the lowest system reward among the four algorithms. Furthermore, the DIL curves are oscillating and unstable, with large fluctuation intervals. In comparison, the FL algorithms can achieve higher system reward, and the FL curves are smoother. This is because in the DIL algorithm, there is no experience sharing between UEs and some UEs need to spend more time exploring and trying non-optimal actions. UEs may choose some bad actions during the exploration process, thus causing large fluctuations in the network performance and the reward value. Compared with the FL algorithm, the FLI algorithm can achieve higher system reward. This is because the parameter initialization from the expert model to newcomers can help them start faster. Also, newcomers may contribute lousy experience to the global model which will interfere with the experience sharing if they fail to learn practical experience before the federated learning process. The initialization can prevent users from contributing lousy experience, so it will help enhance the system performance. This can demonstrate that in addition to the regular role in federated learning, the global model can also play an additional role as an expert model for transfer learning. Finally, it can be observed that our proposed KT-FLUC algorithm can achieve the highest system reward, lowest delay, and highest throughput. This verifies that the proposed knowledge transfer algorithm performs better than copying parameters directly. The combination of federated learning and knowledge transfer can further improve system performance compared with federated learning.

% \begin{figure*}[ht]
%     \centering
%     \subfigure[The average system reward with different UE numbers.]{
%         \includegraphics[width=2.25in]{UE-4.png}
%     }
%     \subfigure[The average GBR traffic delay with different UE numbers.]{
% 	\includegraphics[width=2.25in]{UE-5.png}
%     }
%     \centering
%     \subfigure[The average non-GBR traffic throughput with different UE numbers.]{
%     	\includegraphics[width=2.25in]{UE-6.png}
%     }
%     \caption{The network performance under different number of UEs of FTL, CL and DIL algorithms. (a) The average GBR traffic delay. (b) The average non-GBR traffic throughput. (c) The system reward.}
%     \label{fig11}
% \end{figure*}

In Fig. \ref{fig9}, the average delay of GBR traffic and throughput of non-GBR traffic under different numbers of UEs are shown. With FLI and proposed KT-FLUC algorithms, there is no significant change in delay while the throughput will decrease when the number of UEs increases. This is because GBR traffic has a higher priority than non-GBR traffic, demonstrating that the traffic has been directed more effectively. In summary, the proposed KT-FLUC algorithm can consistently achieve the lowest packet delay and the highest throughput with different UE numbers. When the average UE number is 45, the proposed KT-FLUC algorithm can achieve 24\% lower delay and 43\% higher throughput compared with DIL algorithm. Also, with the proposed knowledge transfer scheme, it can achieve 10\% lower delay and 10\% higher throughput compared with FLI algorithm.

\begin{table}[t]
\caption{Running time of KT-FLUC and FLI.}
\label{table2}
\vspace{-5pt}
\renewcommand{\arraystretch}{1.2}
\centering
\begin{tabular}{|l|l|l|}
\hline
                            & KT-FLUC & FLI     \\ \hline
Federated learning time (s) & 25.0549 & 28.6939 \\ \hline
Knowledge transfer time (s) & 0.0291  & 0.0243  \\ \hline
Other time (s)              & 1.9063  & 1.8299  \\ \hline
Total time (s)              & 26.9903 & 30.5481 \\ \hline
\end{tabular}
\vspace{-10pt}
\end{table}

Table \ref{table2} further compares the running time of the KT-FLUC algorithm and the FLI algorithm. It shows the running time of different modules of the framework over 500 TTIs. In this table, the other time includes all other processes during the simulation, the PRB scheduling, and the data transmission. It can be observed that the knowledge transfer time of FLI is larger than KT-FLUC since the expert model will cause additional computational workload. However, the parallel running method used by KT-FLUC can efficiently reduce the federated learning time and total running time of the framework. Although extra calculation steps are introduced in KT-FLUC, it can still reduce the total running by 10\%.

\begin{figure}[ht]
    \centering
    \subfigure[The average number of UE in the system is 25]{
        \includegraphics[width=8.5cm,height=5.8cm]{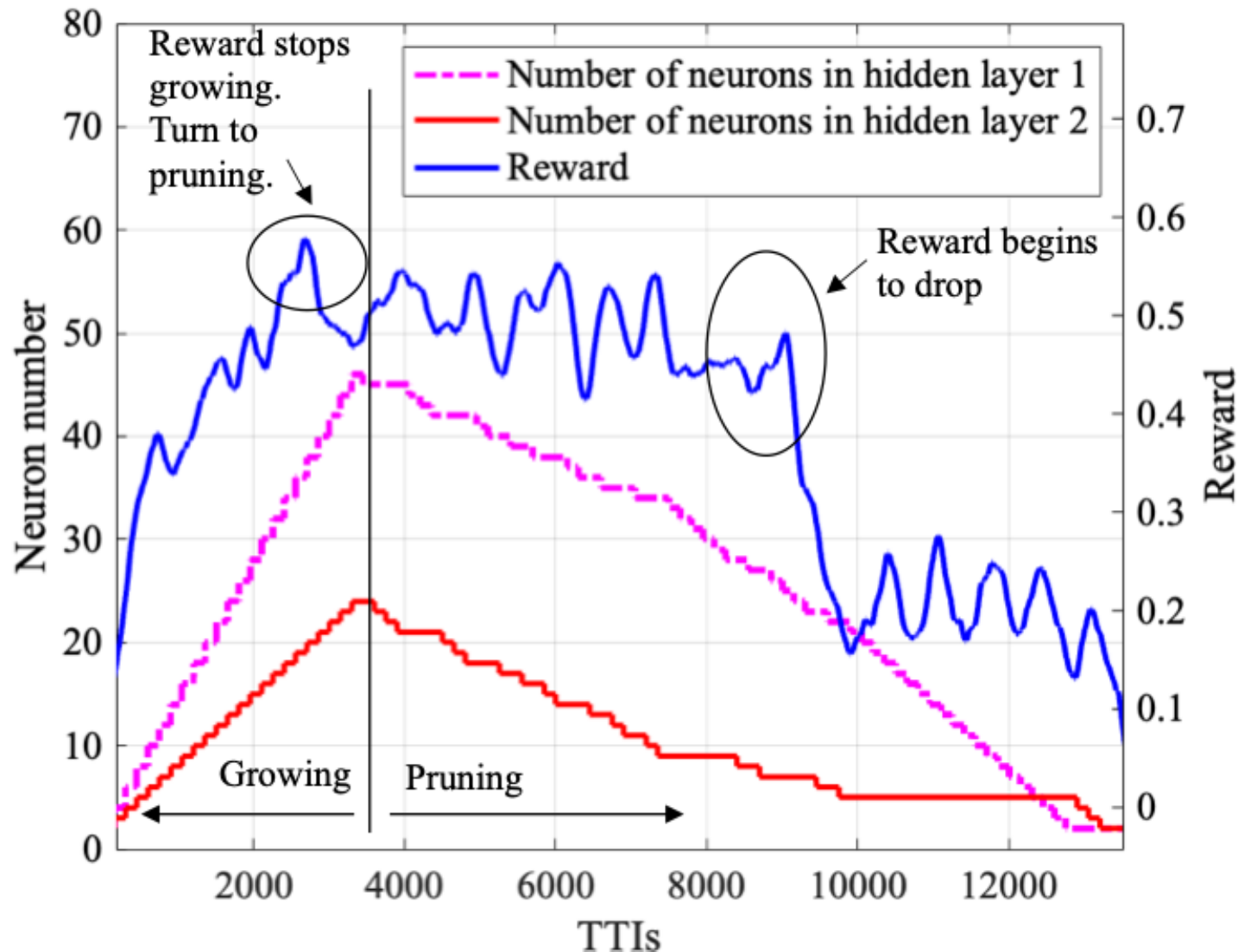}
    }
    \subfigure[The average number of UE in the system is 45]{
	\includegraphics[width=8.5cm,height=5.8cm]{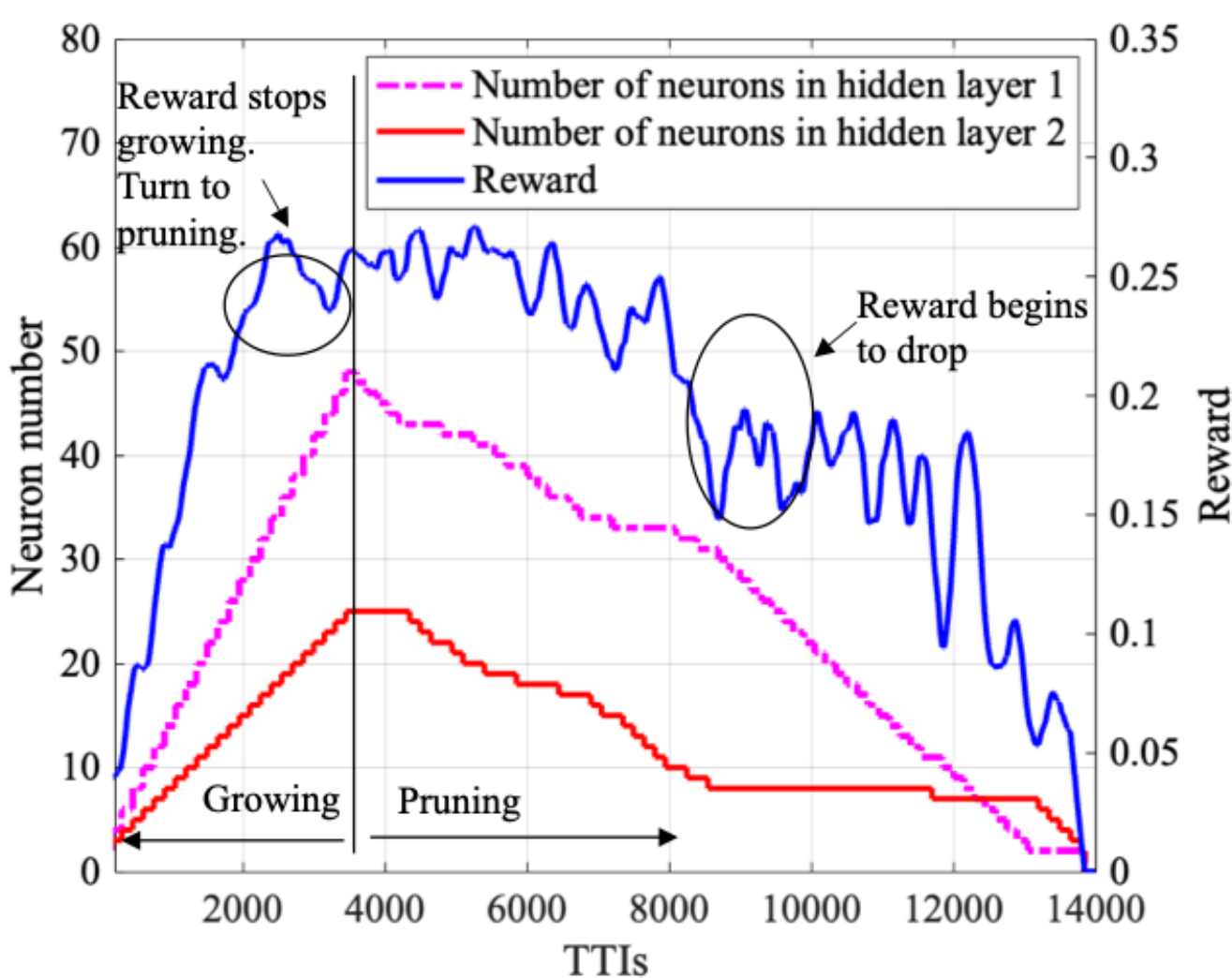}
    }
    \subfigure[The average number of UE in the system is 65]{
    	\includegraphics[width=8.5cm,height=5.8cm]{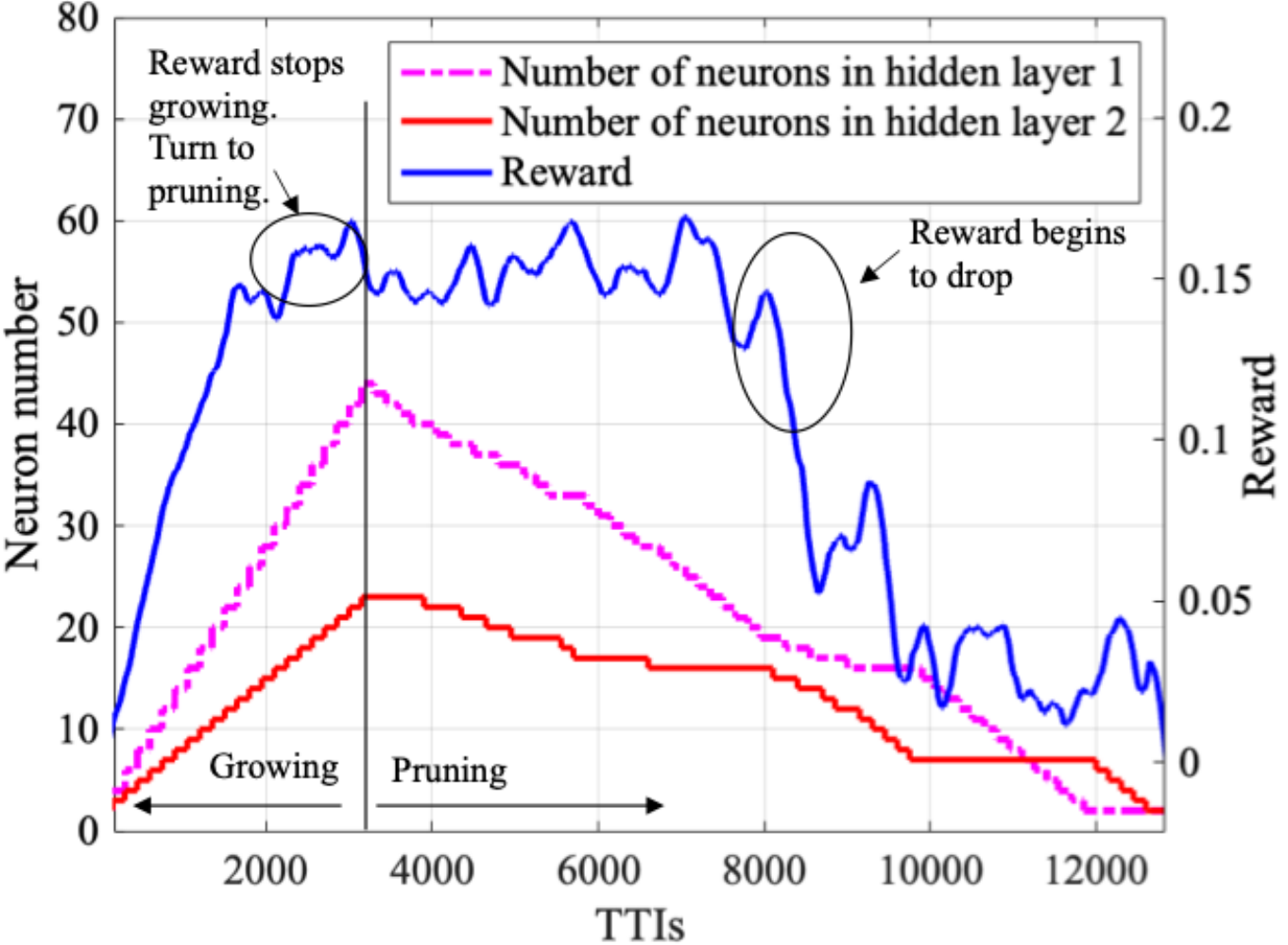}
    }
    \caption{The variation of neuron numbers in hidden layers and system reward during the growing and pruning simulation process under different UE numbers.}
    \label{fig6}
    \vspace{-25pt}
\end{figure}

\subsection{Performance Analysis of Model Compression}

In this section, we analyze the simulation results of the pre-simulation for model compression. We assume the neuron number of the second layer is half of the first layer in the growing phase since the dimension of output layer is much smaller than the dimension of the input layer and 2:1 is a very common assignment of neurons among two hidden layers. The variation of the number of neurons in the hidden layer and the system reward during the pre-simulation is shown in Fig. \ref{fig6}. The numbers of UEs in Fig. \ref{fig6} (a), Fig. \ref{fig6} (b) and Fig. \ref{fig6} (c) are respectively 25, 45 and 65.

We take Fig. \ref{fig6} (a) as an example to further illustrate the growing and pruning process for 32 UEs. In the beginning, the system is in a growing phase, and the neuron number of hidden layers increases with time. As a joint result of the response to increasing model size and model training, the system reward is also growing. Then during the 1800 - 2700 TTIs, a stabilization of the increase in reward is observed which is marked by an
ellipse and arrow in the figure. The stabilization indicates that model training has begun to converge, and the current model size is large enough to support the TS application. So the simulation steps into the pruning process and the neuron numbers of both layers begin to decrease. Fig. \ref{fig6} (b) and Fig. \ref{fig6} (c) show similar processes. The neuron number increases during the growing phase the reward stabilizes and then decreases during the pruning phase.

At the beginning of the pruning process, the reward keeps stable as neurons with weaker competitiveness are pruned. This trend validates the theory that less competitive neurons have less influence on the output and demonstrates the possibility of reducing the number of hidden neurons without affecting the system performance to realize model compression. After about 8000 TTIs, there is a noticeable drop in the reward, meaning the corresponding model size is not large enough to support the given network function. The system performance will be influenced if we continuously compress the system model by pruning neurons. Therefore, by observing the reward during the pruning process, we can find a threshold as the maximum compression achievable without compromising system performance.

\begin{figure}[t]
\centerline{\includegraphics[width=9cm,height=6.3cm]{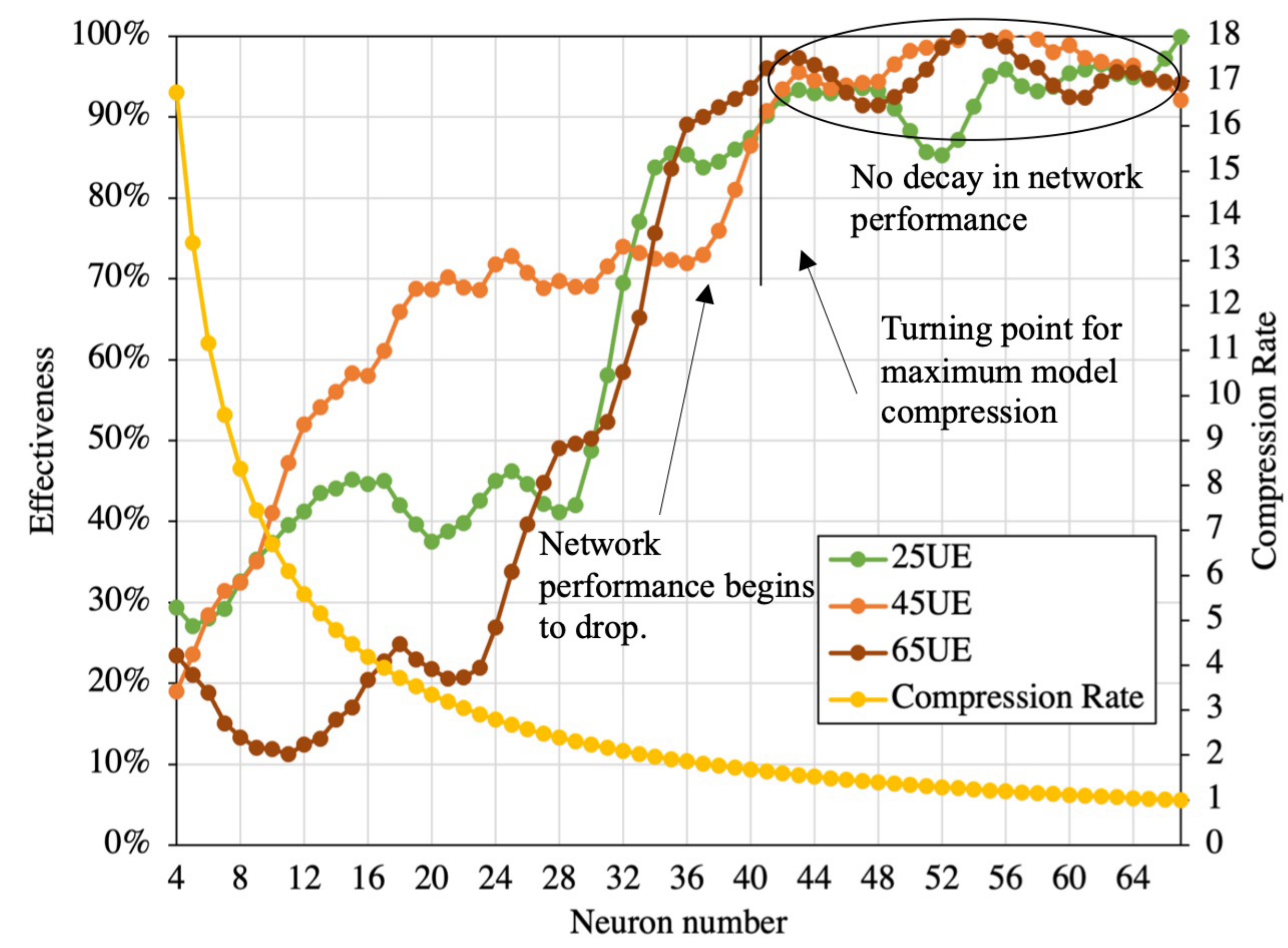}}
\caption{Effectiveness and compression rate of different hidden neurons under different UE numbers.}
\label{fig7}
\vspace{-15pt}
\end{figure}

The further determine the compression threshold value, we defined an indicator called effectiveness by counting the average reward under a given model size and normalizing it with the maximum reward. The effectiveness and model compression rate under different model sizes is shown in Fig. \ref{fig7}. %From Fig. \ref{fig7}, it can be observed that the scenarios with more UEs, the system performance drops at a faster speed compared with scenarios with fewer UEs 
It can be observed that when the neuron number is larger than 40, few decays can be observed in the system performance of all three scenarios with effectiveness above 90\%. On the other hand, when the neuron number is less than 40, the effectiveness of models has declined significantly to varying degrees. So we conclude that 40 neurons is the threshold value for the compressed model size and the corresponding compression rate, 1.5, is the maximum compression rate achievable without compromising performance. In this situation, there will be 80 network parameters need to be submitted during federated learning, and if there are 50 UEs, the size of information to be exchanged is about 8 KB. Considering that at the initial pruning phase, the neurons of two hidden layers are evenly pruned, we keep the two-fold relationship of neuron numbers of two hidden layers and use 14 and 28 neurons at each layer as our model size for following TS simulations.

\vspace{-10pt}
\section{Conclusion}\label{s6}

%\red{It is too long as a conclusion, you can start with "ML is a promising solution for the management of xxx." Read your former papers as reference.}
%TS is an effective technique to distribute traffic flows among UEs. Nowadays, UE-centric TS is attracting more attention than cell-based learning considering the UE-centric 5G access network architecture and the concerns about privacy, latency and efficiency. 
Machine learning is a promising solution for optimizing wireless networks. In this paper, we propose a federated transfer reinforcement learning-enabled UE-centric TS framework with knowledge transfer and model compression. %This framework is composed of three essential techniques, federated learning, transfer learning and model compression. 
The core idea is to use federated learning to aggregate decentralized local models, generate a robust global model, and use knowledge transfer to initialize newcomers. The simulation results show that the proposed framework achieves better convergence, stability, higher system throughput and lower delay compared with independent UE-centric TS and cell-centric TS. Considering these benefits, our proposed framework is not limited to TS. It can also be applied to other distributed wireless applications, such as edge caching and sleep control. 
In the future, we plan to extend UE-centric wireless networks to other applications.

% or


% \appendices

% % you can choose not to have a title for an appendix
% % if you want by leaving the argument blank
% \section{}

\vspace{-10pt}
\section*{Acknowledgment}
 This work has been supported by MITACS and Ericsson Canada, and NSERC Collaborative Research and Training Experience Program (CREATE) under Grant 497981.

\ifCLASSOPTIONcaptionsoff
  \newpage
\fi

\end{document}